\documentclass[fleqn]{annalen}
\usepackage{graphics}
\usepackage{epsf}
\usepackage{epsfig}
\begin{document}
%%%%%%%%%%%%%%%%%%%%%%%%%%%%%%%%%%%%%%%%%%%%%%%%%%%%%%%%%%%%%%%%%%%%%%%%%%%%%%
%%%%%%%% the following new commands will be completed by the publisher %%%%%%%%
%%%%%%%%%%%%%%%%%%%%%%%%%%%%%%%%%%%%%%%%%%%%%%%%%%%%%%%%%%%%%%%%%%%%%%%%%%%%%%
\newcommand{\volume}{9}              %sets current volume,
\newcommand{\xyear}{2000}            %sets year in header
\newcommand{\issue}{1}               %sets current issue,
\newcommand{\recdate}{28 April 2000}    %sets received date,
\newcommand{\revdate}{dd.mm.yyyy}    %sets revised date,
\newcommand{\revnum}{0}              %number of revisions,
\newcommand{\accdate}{06 July 2000}    %sets accepted date,
\newcommand{\coeditor}{F.W. Hehl}           %sets (co)editor,
\newcommand{\firstpage}{1}           %first page number,
\newcommand{\lastpage}{33}            %last page number,
\setcounter{page}{\firstpage}        %sets page counter to first page number
%%%%%%%%%%%%%%%%%%%%%%%%%%%%%%%%%%%%%%%%%%%%%%%%%%%%%%%%%%%%%%%%%%%%%%%%%%%%%%
\newcommand{\keywords}{cosmic phase transitions, deconfinement, inflation}
\newcommand{\PACS}{04.40,-b, 24.85, 98.80.Cq, 64.60.Qb}
\newcommand{\shorttitle}{B. K\"ampfer, Cosmic Phase Transitions
} %% sets the header on oddpage
\title{Cosmic Phase Transitions}
\author{B. K\"ampfer$^{1}$}
\newcommand{\address}
  {$^{1}$Forschungszentrum Rossendorf, Postfach 510119,
  01314 Dresden, Germany}
\newcommand{\email}{\tt kaempfer@fz-rossendorf.de}
\maketitle
%%%%%%%%%%%%%%%%%%%%%%%%%%%%%%%%%%%%%%%%%%%%%%%%%%%%%%%%%%%%%%%%%%%%%%%%%%%%%
\begin{abstract}
The sequence of phase transitions during the hot history of the universe
is followed within a phenomenological framework.
Particular emphasis is put on the QCD confinement transition, which
is at reach under earth laboratory conditions.
A tepid inflationary scenario on the GUT scale
with bubble growth at moderate supercooling is discussed.
\end{abstract}

%%%%%%%%%%%%%%%%%%%%%%%%%%%%%%%%%%%%%%%%%%%%%%%%%%%%%%%%%%%%%%%%%%%%%%%%%%%%%

\section{Introduction}

The standard cosmology teaches that the universe is expanding.
This implies a steady change of the state of matter. Particularly interesting
are phase transitions where the structure and the relevant degrees of freedom
change. Probably the most drastic transition was the formation of the
early universe. There is a number of ideas how did this happen,
including the creation from a quantum state (cf.\ \cite{Kiefer}
and further references there).
Since the extrapolation of our quantum field theoretical models to high
temperature and density points to a maximum temperature of
$T \sim 10^{16}$ GeV, where thermal equilibrium between various particles
can be maintained, one can guess that the thermal history starts
at such a temperature scale and corresponding world age of
$10^{-37}$ sec. Immediately afterwards one conjectures that a symmetry
breaking phase transition happens. At a scale of $T \sim 10^{16}$ GeV,
or slightly less,
the extrapolated coupling strengths of the strong, weak and electromagnetic
interactions merge, and one generally believes that the behavior of matter
is to be described within the framework of a grand unified theory (GUT).
The very structure of GUT is still matter of debate, and also the relation
to supersymmetric and string theories is not yet settled down.
Nevertheless, inflation, i.e. a stage of accelerated expansion and a huge
blowing off of the space is an important scenario, which allows to
understand some of the presently observed properties of the universe.

The subsequent evolution appears quite unspectacular for a long time span:
expansion means cooling,
and particles with masses in the order of 40\% of the corresponding
temperature disappear due to annihilation.

The next interesting stage is the symmetry breaking on the electroweak
scale, i.e. $T \sim 100$ GeV. Cooling further, at $T \sim 160$ MeV
the quarks and gluons, roaming up to this stage freely, become
bound in hadrons. This is the confinement transition, which might be called
the hadrosynthesis stage.

Once the temperature falls below 1 MeV the nucleosynthesis starts,
creating the light primordial elements.
In nuclear matter, a liquid-gas phase transition is conjectured since a
long time (cf.\ \cite{our_liquid_gas}), and experimental hints have
been verified \cite{liquid_gas}. However, in the cosmic evolution
this transition does not play a role since it is related
to large baryon density.

From such a point of view the history of the early universe can be considered
as sequence of phase transitions. Indeed, based on quite general
models, one can develop a scheme to combine the dynamics of phase transitions
with the cosmic expansion, as envisaged in \cite{Boyanowski}.
Here we follow, however, a different avenue and consider the specific
features of individual transitions.

Among the above mentioned phase transitions the confinement transition
from quarks and gluons to hadrons has been elaborated by far in most details.
This is because it is believed to be reproduceable in the laboratory.
Indeed, for colliding nuclei at sufficiently high energies,
the relevant degrees of freedom
% of strongly interacting matter 
should be represented by partons,
i.e. point-like structures.
%for a short instant of time.
The presently terminating first
round of heavy-ion experiments at CERN accumulated such a wealth of data
which give rise to the interpretation that in central nuclear collisions
a thermalized state of deliberated quarks and gluons has been created.
Indeed, on February 9, 2000 a press release of CERN announced
``compelling evidence for the existence of a new state of
matter in which quarks, instead of being bound up into more complex particles
such as protons and neutrons, are liberated to roam freely, $\cdots$
this state must have existed at about 10 microseconds after the Big Bang''
\cite{CERN}.

In accordance with this new information we will focus here on the
confinement transition after the Big Bang. We will compare schematically
the confinement dynamics of the Big Bang
with that of the Little Bang (i.e. relativistic heavy-ion
collisions).
By using the known details of the transition
we demonstrate the interplay of local matter properties and global
evolution dynamics. More specifically, we analyze
how drives the local matter state
the global expansion, and vice versa, how the expansion causes
cooling and transformation of matter forms into each other.

\begin{figure}
 \psfig{file=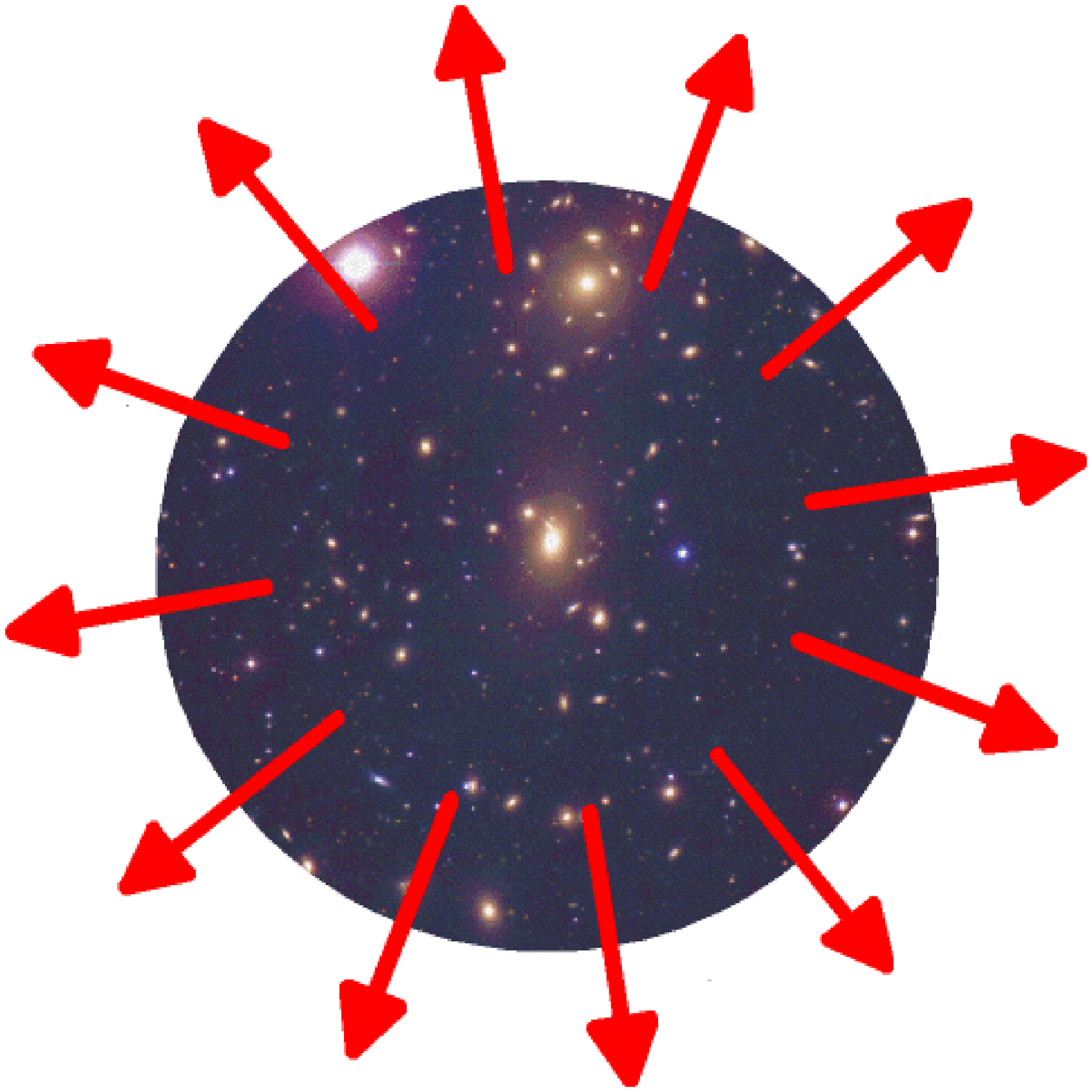,width=5.9cm,angle=-0}
 \hfill
 \psfig{file=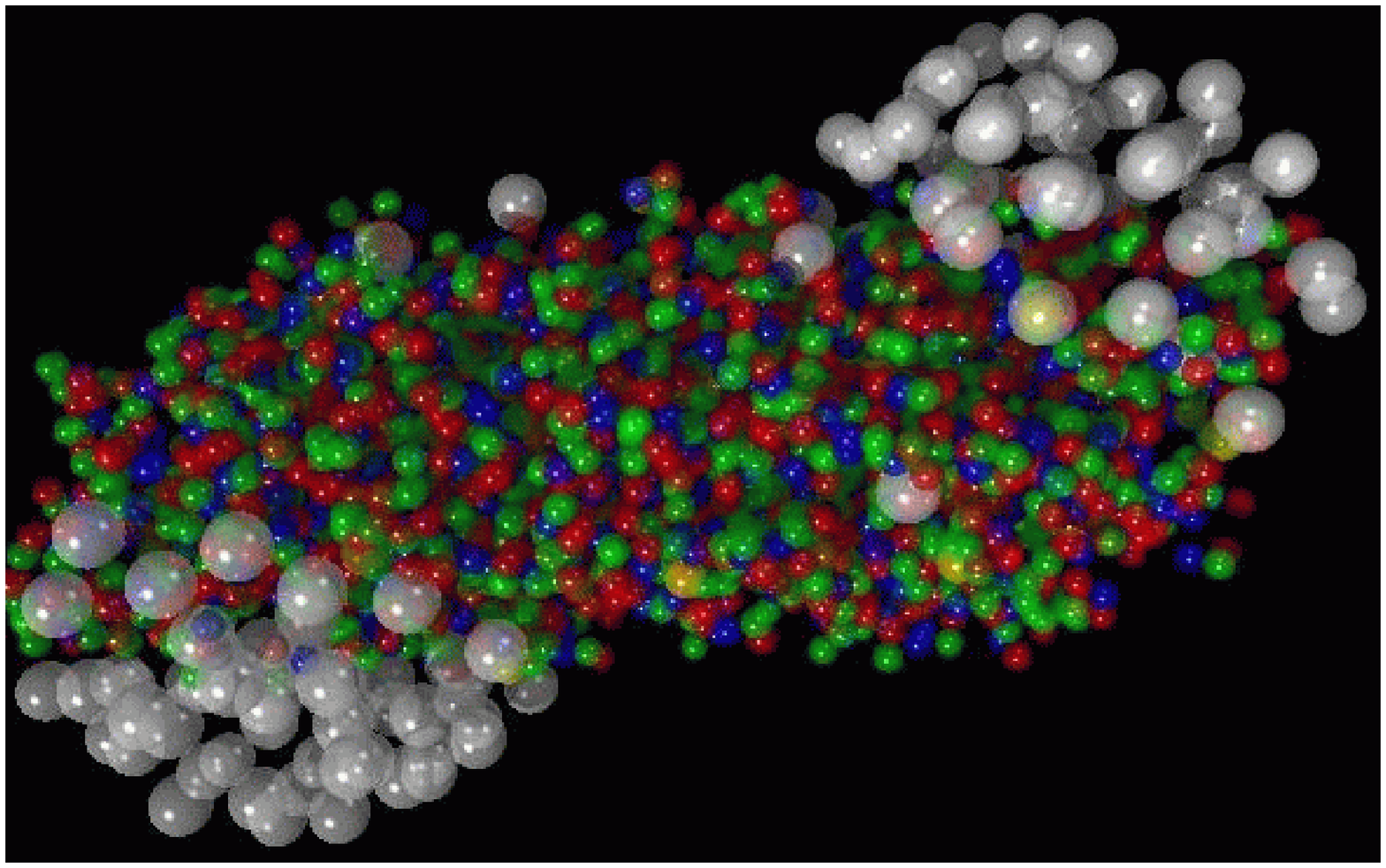,width=3.9cm,angle=90}
 \caption{Expanding matter: The left panel displays a view on
 a group of remote galaxies as seen by the Hubble telescope
 (source of the underlying picture: NASA picture gallery in {\tt www}
 \cite{NASA_1}),
 while the right panel shows the distribution of hadrons in
 a semi-peripheral collision of heavy nuclei a few fm/c after the first
 touch (source: CERN press release in {\tt www} \cite{CERN}).}
 \label{expanding_matter}
\end{figure}

There is some apparent similarity in cosmology
and heavy-ion physics,  which we try to visualize
in Fig.~\ref{expanding_matter}. From present observations
of remote objects in the universe,
we look somewhat into the past. Combining observational
facts, like the distribution and the red shift of galaxies, as seen
in the left panel of Fig.~\ref{expanding_matter},
one can develop a picture of the early stages of the universe.
The situation in heavy-ion physics is very similar to this.
Observing the created
hadrons in a very late stage, one tries to extrapolate back to the
hottest and densest stages. The artist's view of a semi-peripheral
collision of heavy-ions, displayed in the right panel of
Fig.~\ref{expanding_matter}, shows a stage where the matter is converted into
hadrons which still interact for a while before disassembling.
However, there is a chance in heavy-ion collisions to receive a direct
signal of the very early stage: the mean free path of real and virtual
photons \cite{direct_probes}
is large in comparison with the system size.
For such direct probes the problem, however, arises
to disentangle the primordial signal from a huge background
\cite{Gallmeister_2}.

The present survey is restricted to a phenomenological framework.
That means, we do not go into a microscopic theory of phase transitions,
rather describing them only in terms of the equation of state.
We restrict ourselves to the standard model and the grand unifying theory
models of the strong and electroweak interactions. Only in a few
exceptional cases we quote interesting effects occurring beyond the
standard model of particle physics within supersymmetric or string-theoretical
models.

The paper is organized as follows.
In section 2 the phenomenological framework is presented resulting in
Friedmann's equations for the Big Bang and the corresponding equations
of matter in the Little Bang.
Section 3 deals with quantum chromodynamics (QCD), quarks and gluons
and their equation of state. Section 4 exercises a simple way to
construct first-order phase transitions. The basic issues of nucleation
theory for handling the phase transformation dynamics are presented in
section 5.
Section 6 is devoted to details of the cosmic confinement transition.
A short remark concerning the present status of knowledge on the
electroweak transition is made in section 7.
In section 8 we deal with an attempt of a phenomenological realization
of an inflationary scenario. The summary can be found in section 9.

The material represented is based on {\tt www} scans till end of April 2000.

\section{Phenomenological framework}

Our phenomenological considerations are based on two cornerstones,
namely the description of\\
(i) local matter properties within the framework of
thermodynamics and hydrodyna\-mics, and\\
(ii) global cosmic expansion by
Einstein's equation and the cosmological principle.\\
We are now going to present these issues and define the propositions.

\subsection{Einstein equations}

The Einstein equations can be written as
\begin{eqnarray}
R_{ij} - \frac 12 g_{ij} \hat R = \kappa \hat T_{ij},
\quad\quad
\kappa = \frac{8 \pi G_N}{c^4},
\label{Einstein}
\end{eqnarray}
which can be interpreted as gravity, described by geometry or curvature
of the space-time on the left-hand-side, being caused by matter,
which in turn is
described by the phenomenological energy momentum tensor,
$\hat T_{ij}$, on the right-hand-side. More specifically,
$R_{ij}$ is the Ricci tensor, $\hat R = R_{ij} g^{ij}$
the curvature scalar, and $g_{ij}$
the metric. The Newtonian constant $G_N$ serves as
coupling strength of geometry to its source, i.e. to matter.
In the following we employ mainly units from particle physics, i.e.
$\hbar = c = k_B =1$ for the Planck quantum divided by $2\pi$, and
velocity of light, and Boltzmann's constant, respectively.\footnote{
The conversion constant
$\hbar c = 197.327053$ MeV$\cdot$fm is useful for relating
length scales and energy scales and for translations into
other unit systems as well.}
The energy-momentum tensor can be split into a vacuum part,
$T_{ij}^{\rm vac}$, and a matter part, $T_{ij}$, accounting for
excitations above the vacuum:
\begin{eqnarray}
\hat T_{ij} = T_{ij}^{\rm vac} + T_{ij},
\quad\quad
T_{ij}^{\rm vac} = (e^{\rm vac} + p^{\rm vac}) u_i u_j
- p^{\rm vac} g_{ij},
\end{eqnarray}
where $e^{\rm vac} = - p^{\rm vac}$ ensures the local 
Lorentz invariance of the vacuum.
The presently observed accelerated expansion of the universe
\cite{acc_exp} could be caused a dominating vacuum energy
$e^{\rm vac} \approx
5 \times 10^{-30} \mbox{g} \cdot \mbox{cm}^{-3} \approx
0.6 \, e^{\rm crit} \approx
10^{-123} M_{\rm Pl}^4$ \cite{e_vac}.
Here, $M_{\rm Pl} = 1.22 \times 10^{19}$ GeV = $ 2.17 \times 10^{-5}$ g
stands for the
Planck mass defined by $M_{\rm Pl} = \sqrt{\hbar c / G_N}$.
The vacuum energy is the missing link to add up all energy
forms, including the substantial part of dark matter \cite{dark_matter},
to an amount corresponding to a flat universe with
$e = e^{\rm crit}$, which is ``predicted'' by inflationary scenarios.
The present critical energy density is
$h^2 1.88 \times 10^{-29}$ g cm${}^{-3}$
with $h \approx 0.68$ \cite{h_0}.

From the Friedmann equations (see (\ref{R2dot}) below) a less
negative ``vacuum'' pressure of $p^{\rm vac} \le - \frac 13 e^{\rm vac}$
is enough to drive an accelerated expansion.
The present vacuum energy could be generated by evolving fields,
e.g.\ the tracker fields \cite{tracker}. This would cause a noticeable
uncertainty in extrapolating back to the early history of the universe,
because the relation of the vacuum energy density to the energy
density of other matter forms is rather unsettled.

If the vacuum energy, sometimes also dubbed quintessence \cite{quintessence},
would be constant, then it can comfortably be included in Einstein's famous
cosmological constant $\Lambda = e^{\rm vac} \kappa = const$,
and the Einstein equations would read
\begin{eqnarray}
R_{ij} - \frac 12 g_{ij} \hat R - \Lambda g_{ij} = \kappa T_{ij}.
\end{eqnarray}
In such a case, the vacuum energy can be neglected at early times,
i.e.\ $e^{\rm vac} \ll e$.
We do not touch upon the questions on the origin of the cosmological
constant \cite{Weinberg} or why the vacuum energy is becoming
operative just now, but we assume in the following
that the present vacuum energy
can be neglected at early times in comparison with other energy
contributions. We mention in advance that in certain cosmic
stages appropriate vacua are included.

\subsection{Cosmological principle and Robertson-Walker metric}

The cosmological principle requires homogeneity and isotropy
resulting in a simple space-time symmetry according to SO(4), or
E(3), or SO(3,1) symmetry groups\footnote{
We do not consider the Kantowski class which has no local Minkowski limit.}
and the Robertson-Walker metric (e.g.\ cf.\ \cite{cosmology_book})
\begin{eqnarray}
ds^2 = dt^2 -
R(t)^2 \left( \frac{dr^2}{1 - k r^2} + r^2 d \Omega^2 \right),
\label{RW_metric}
\end{eqnarray}
where the only dynamical quantity is the scale factor $R(t)$.
The parameter $k = \pm 1, 0$ determines whether the universe has
a closed, open, or flat geometry. For early cosmology and our purposes
it is sufficient to consider the case $k = 0$.

\begin{figure}
 \vskip -.03cm
 ~\center
 \psfig{file=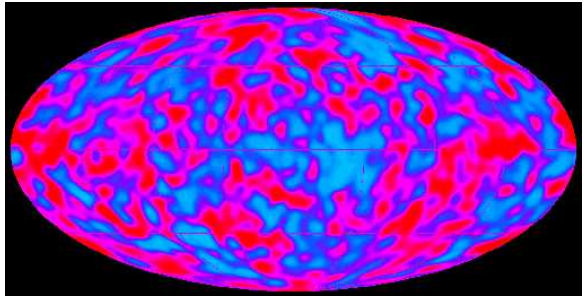,width=3.9cm,angle=-90}
 \caption{A contour plot of the temperature distribution in the universe
 at last rescattering of photons as seen by the cosmic background explorer
 satellite COBE (source: NASA picture gallery in {\tt www} \cite{NASA_2}).} 
 \label{COBE}
\end{figure}

The present microwave background radiation corresponds to a blackbody
radiation with temperature of $T = 2.7277$ K with relative fluctuations
less than $10^{-4}$. The cosmic background explorer COBE quantifies
these fluctuations, as shown in Fig.~\ref{COBE}.
This COBE map may be understood as a snapshot of the universe at
the age of 400,000 years. While the tiny fluctuations have been amplified
during the last 14 Gigayears to the present structures, like galaxies and
clusters, the very early stages of the universe
can be assumed to be very smooth.

\subsection{Perfect fluid}

In the following we describe matter by the energy-momentum tensor
\begin{eqnarray}
T_{ij} = e u_i u_j + q_i u_j + u_i q_j + p_{ij},
\label{T_ij}
\end{eqnarray}
where $e$ is the energy density.
The stress part, $p_{ij}$, is orthogonal to the four-velocity $u_i$
of matter,
i.e. $p_{ij} u^i = 0$, as is
the energy flow $q_i$.
Within this framework,
matter is described by a fluid thus covering special cases
as dust, real fluid, gas, and vacuum.
In (\ref{T_ij}) it is assumed that one unique velocity field
is common to all subcomponents of matter.
There are two possible gauges of the velocity:
either according to Landau, with $u_i$ parallel to energy flow, or
according to Eckart, with $u_i$ parallel to baryon flow.
Since we neglect in most examples the baryon charge, the Eckart gauge
is applied.

In the following we mainly consider non-dissipative media,
i.e. heat conductivity and shear viscosity and possible other
gradient terms with non-standard dissipative effects
\cite{non_standard_hydro} are not included.
The only dissipative effect we consider in subsection \ref{viscosity}
is the volume viscosity parameterized by the coefficient $\hat \xi$
entering the stress via
\begin{eqnarray}
p_{ij} = (p - \hat \xi u^i{}_{;i})(- g_{ij} + u_i u_j), \quad\quad
u^i u_i = + 1.
\label{p_ij}
\end{eqnarray}
The entropy current then reads
\begin{eqnarray}
s_i = s u_i % + T^{-1} q_i
\label{entropy_current}
\end{eqnarray}
and has to satisfy $s^i{}_{;i} \ge 0$; here
$s$ is the entropy density;
% and $T$ stand for the temperature;
the semicolon, ``;'', denotes the covariant derivative.
(\ref{T_ij}, \ref{p_ij}) with $\hat \xi = 0$ define the
perfect fluid.

\subsection{Equations of motion of perfect fluid in Big Bang}

Discarding any dissipative effect we now present the equations of
motion of matter and geometry for the Big Bang.
Since the cosmic matter consists of several components, labeled by
$\alpha$, we take into account
\begin{eqnarray}
e = \sum_\alpha e_\alpha, \quad\quad
p = \sum_\alpha p_\alpha.
\end{eqnarray}

\subsubsection{Robertson-Walker metric}

The four velocity in comoving coordinates is
$ u_i = \delta_i^0$ yielding with the metric (\ref{RW_metric})
\begin{eqnarray}
u^i{}_{;i} = 3 \frac{\dot R}{R}.
\end{eqnarray}

\subsubsection{Projection of the contracted Bianchi identity}

The projection of the contracted Bianchi identity
$u_i T^{ij}{}_{;j} = 0$ yields
\begin{eqnarray}
\dot e + 3 \frac{\dot R}{R} (e + p) = 0
\end{eqnarray}
for the perfect fluid.

\subsubsection{Current conservation}

If there are $\beta$ conserved currents, $(n_\beta u^i)_{;i} = 0$,
the conservation of the corresponding charges reads
\begin{eqnarray}
\dot n_\beta + 3 \frac{\dot R}{R} n_\beta = 0
\quad \to \quad n_\beta R^3 = const.
\end{eqnarray}
The quantities $n_\beta$ can represent the baryon density, or the
electric charge density etc.

\subsubsection{Einstein's equations}

For $\Lambda = 0$ the Einstein equations (\ref{Einstein}) read
\begin{eqnarray}
R_{ij} - \frac 12 g_{ij} \hat R = 3 {\cal C}^2 \hat T_{ij},
\quad\quad
{\cal C} \equiv M_{\rm Pl}^{-1} \sqrt{\frac{8 \pi}{3}}.
\label{eq.12}
\end{eqnarray}
A consideration of $T_{11} - \hat R T_{00}$ and the corresponding
left hand side of (\ref{eq.12}) yields
%yield from the combination of the 11 component minus the
%$R \times$ 00 component
\begin{eqnarray}
\dot R = \sqrt{{\cal C}^2 R^2 e - k}.
\end{eqnarray}
For $k = 0$ these equations can be combined to result in
\begin{eqnarray}
\dot R & = & {\cal C} R \sqrt{e},
\label{Friedmann1}\\
\dot e & = & - 3 {\cal C} (e + p) \sqrt{e}.
\label{Friedmann2}
\end{eqnarray}
(Notice that for early cosmology the distinction of
$k = \pm 1$ or $0$ is usually not important.)
To integrate the latter equation, an equation of state, $p(e)$,
is needed. It should be emphasized that the derivation above
does not invoke entropy conservation.

The $i = 0$, $j = 0$ component of the Einstein equations
(\ref{eq.12}) reads
\begin{eqnarray}
\ddot R = - \frac 12 {\cal C}^2 R (e + 3 p).
\label{R2dot}
\end{eqnarray}
This shows that $e + 3 p > 0$ results in $\ddot R < 0$,
i.e., a decelerated expansion, while
$e + 3 p < 0$ means $\ddot R > 0$, i.e., an accelerated expansion.

\subsection{Equations of motion for the Little Bang}

\begin{figure}
 \vskip -.01cm
 \psfig{file=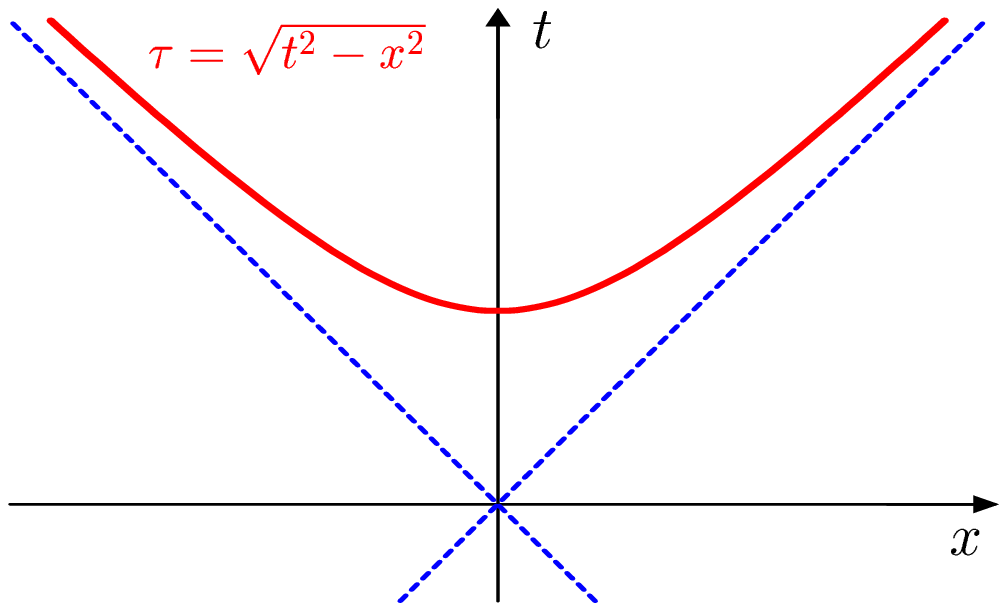,width=6.1cm,angle=-0}
 \hfill
 \psfig{file=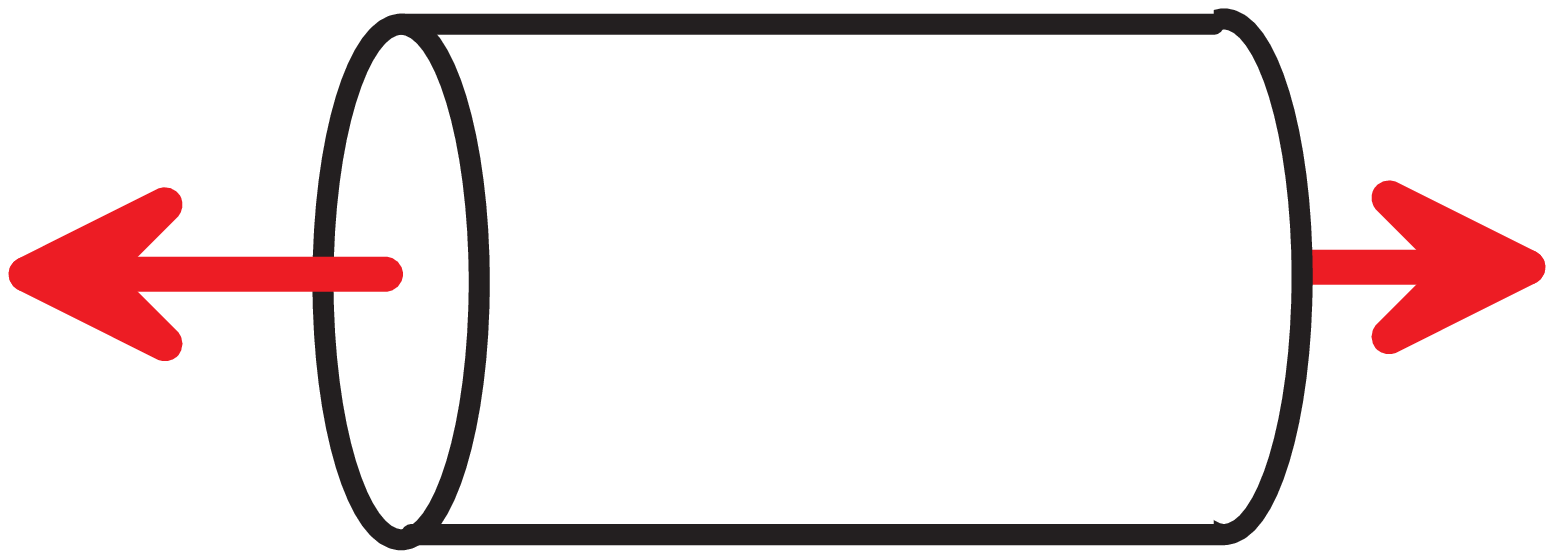,width=6.1cm,angle=-0}
 \caption{Symmetry of a high-energy heavy-ion collision.
 Left panel: Incoming nuclei move nearly on the light cone (dashed lines);
 the thermalized matter has constant energy density and baryon density
 on the hyperboloid of constant proper time $\tau$.
 Right panel: Before the onset of transverse expansion the matter
 expands homogeneously in longitudinal direction.}
 \label{proper_time}
\end{figure}

Various dynamical models \cite{Geiger}
suggest that in a very high-energetic collision
of nuclei the initial conditions on proper time hypersurfaces
(see left panel of Fig.~\ref{proper_time})
$\tau = \sqrt{t^2 - z^2} = const$ are constant. In suitable coordinates,
namely the proper time
$\tau$ and the rapidity $Y = \mbox{atanh} (z/t)$,
this looks like a homogeneously longitudinally expanding fire cylinder
(see right panel of Fig.~\ref{proper_time}).
We neglect the onset of transversal expansion, i.e.,
$v_x = v_y = 0$. Then the motion is called
scale invariant or boost invariant expansion \cite{Bjorken},
which is described by the four-velocity
$u^i = \gamma (1,0,0,v_z)$ with $v_z = z/t$ and
$\gamma = (1 - v_z^2)^{-1/2}$.

The corresponding equations of motion for the perfect fluid then read
\begin{eqnarray}
\dot e = - \frac{1}{\tau} (e + p)
\label{e_dot}
\end{eqnarray}
from $u_i T^{ij}{}_{;j} = 0$ (a dot means here
derivative with respect to $\tau$), and
\begin{eqnarray}
n_\beta \tau = const
\end{eqnarray}
from $(n_\beta u^i)_{;i} = 0$. To integrate (\ref{e_dot}) again the equation
of state $p(e)$ is needed.

Employing a simple equation of state like $p = a T^4$ with
$a \sim {\cal O}(1)$ and the
thermodynamic relations from the next subsection, the dynamical time
scales in Little Bang and Big Bang can be estimated as
\begin{eqnarray}
\mbox{Little \, Bang:} & \frac{e + p}{\dot e} \sim \tau \sim \tau_1
\sim 1 \, \mbox{fm/c}\\
\mbox{Big \, Bang:} & \frac{e + p}{\dot e} \sim \frac{M_{\rm Pl}}{\sqrt{e}}
\sim 10^{19} \, \mbox{fm/c},
\label{time_scale}
\end{eqnarray}
where 1 fm/c = $3 \times 10^{-24}$ sec, and we used a temperature scale of
$T \sim 0.2$ GeV $\sim$ $2 \times 10^{12}$ K.
Eq.~(\ref{time_scale}) shows that
(i) higher energy densities cause faster expansion, and
(ii) the time scale of the Big Bang is determined by the Planck mass.

\subsection{Thermodynamics}

To complete the list of propositions let us recall the thermodynamic
relations for a single-component and single-phase medium:
\begin{eqnarray}
\mbox{thermodynamic \, potential:} & p(T,\mu), \\
\mbox{Gibbs relation:} & e + p - T s = \mu n, \\
\mbox{Euler relations:} & s = \frac{\partial p}{\partial T}, \quad
n = \frac{\partial p}{\partial \mu}, \label{Euler}
\end{eqnarray}
where $\mu$ is the chemical potential and $n$ the related conserved
charge density.

For the early universe, one estimates from present observations
and from Big Bang nucleosynthesis calculations,
$\frac{n_B}{n_\gamma} \vert_{t_*} \sim 10^{-10}$
(here $n_B$ is the baryon density and $n_\gamma$ the photon density),
via entropy conservation a baryo-chemical potential
$\frac{\mu_B}{T} \sim 10^{-9}$. Therefore, for many purposes one
can neglect effects of a finite baryon charge, unless one is dealing
with the evolution of the baryon density itself.
The estimates for the lepto-chemical potentials are not on such firm
grounds due to the non-observability of the neutrino background;
but one usually assumes similarly small lepto-chemical potentials.

From the quantum statistics of ideal gases one finds the pressure
as a thermodynamic potential
\begin{eqnarray}
p & = & \sum_s \pm \frac{d_s}{2\pi^2} T^4
\int_0^\infty dx \, x^2 \ln \left[ 1 \pm \exp \left\{
\frac{\mu_s}{T} - \sqrt{\left(\frac{m_s}{T} \right)^2 + x^2}
\right\} \right] \label{pressure} \\
& \approx &
\sum_{m_s \ll T} d_s \frac{\pi^2}{90} T^4 = a T^4,
\label{p_statistic}
\end{eqnarray}
where $s$ labels the particle species, $d_s$ their effective degeneracies
(being the number of degrees of freedom for each boson, and
$\frac 78$ $\times$ the number of degrees of freedom for each fermion,
respectively); the upper (lower) sign in (\ref{pressure}) is for
fermions (bosons).
The pressure is dominated by those particle species with masses
$m_s$ lower than the temperature $T$.
This toy model equation of state gives the time evolution of the
temperature and entropy density
\begin{eqnarray}
\mbox{Little \, Bang:} &
T =  T_1 \left( \frac{\tau_1}{\tau} \right)^{1/3}, &
s \tau  =  const,\\
\mbox{Big \, Bang:} &
T = (2 \sqrt{3 a} {\cal C} t)^{- 1/2}, &
s R^3 = const,
\end{eqnarray}
where $T_1$ is the maximum temperature at which the system is thermalized
and $\tau_1$ the corresponding time.
Clearly, the comoving entropies are constant.

\begin{figure}
 Little Bang:
 \vskip -.01cm
 \psfig{file=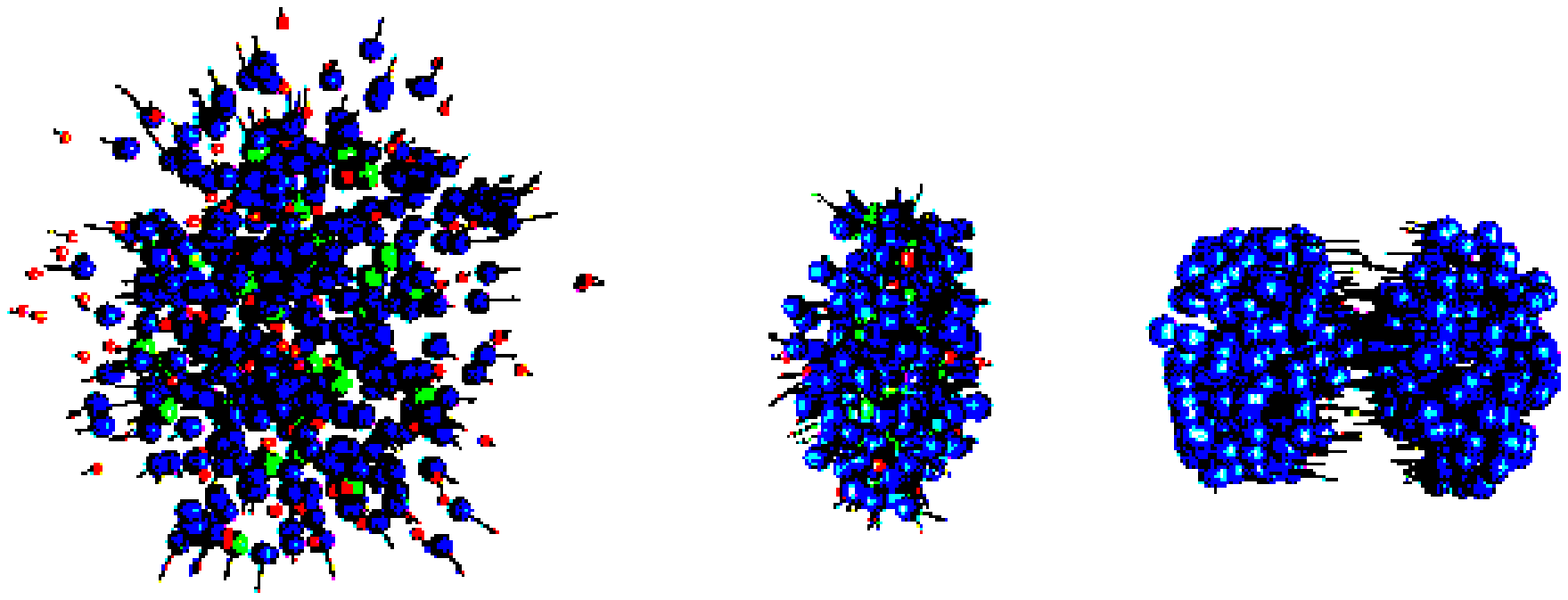,width=3.cm,angle=90}
 \hfill
 \psfig{file=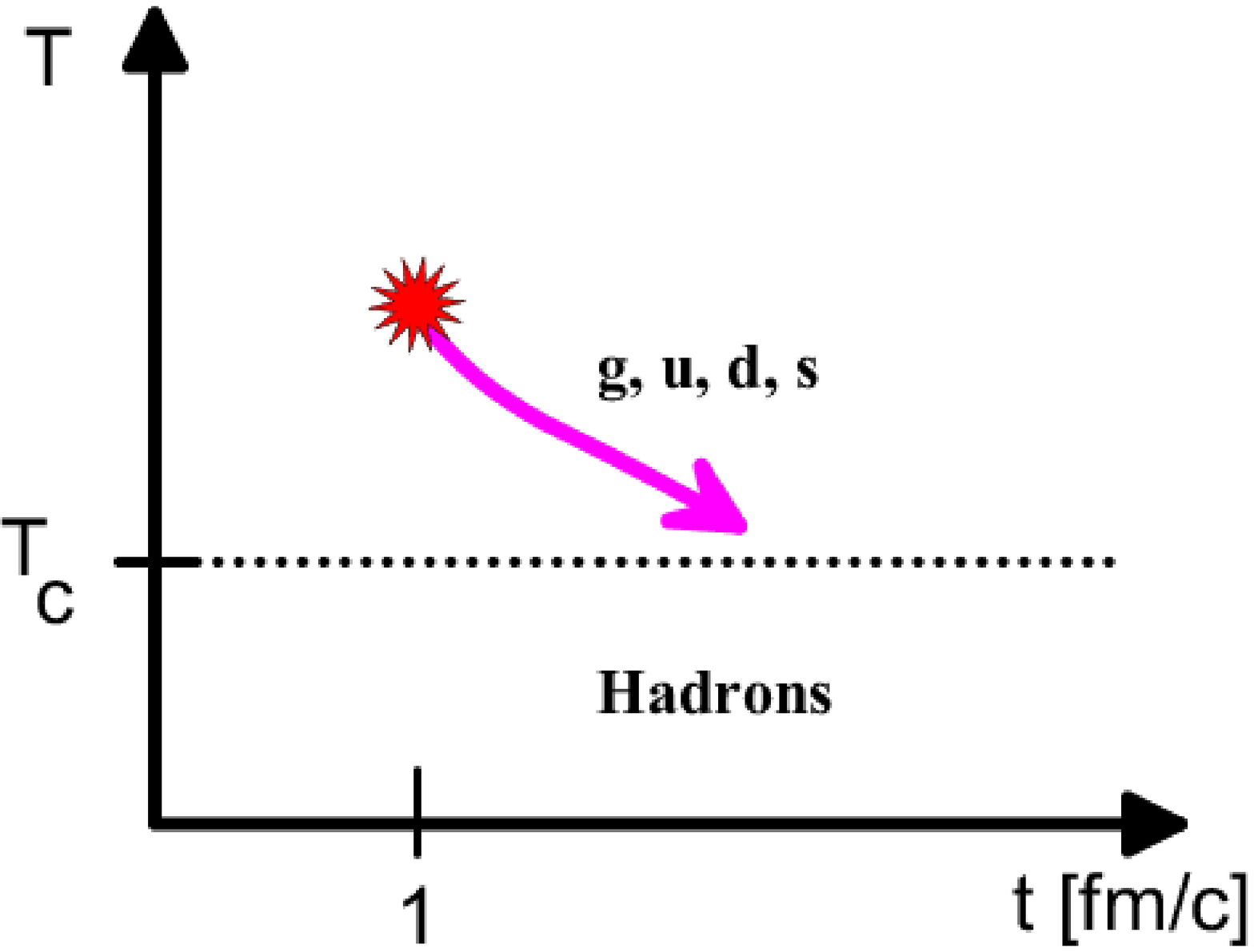,width=4.9cm,angle=-0}
 \caption{Left three panels: An artist's view on a central heavy-ion collision
 (right: the approach stage before first touch,
 middle: highly compressed
 strongly interacting matter is formed; in the center of the blob
 quarks and gluons are the relevant degrees of freedom,
 left: the system has completely transformed to hadrons and expands.
 Pictures by S. Bass exploiting a transport model).
 Right panel: A sketch of the temperature evolution till confinement.}
 \label{Little_Bang}
\end{figure}
\begin{figure}
 Big Bang:
 \vskip -.06cm
 \psfig{file=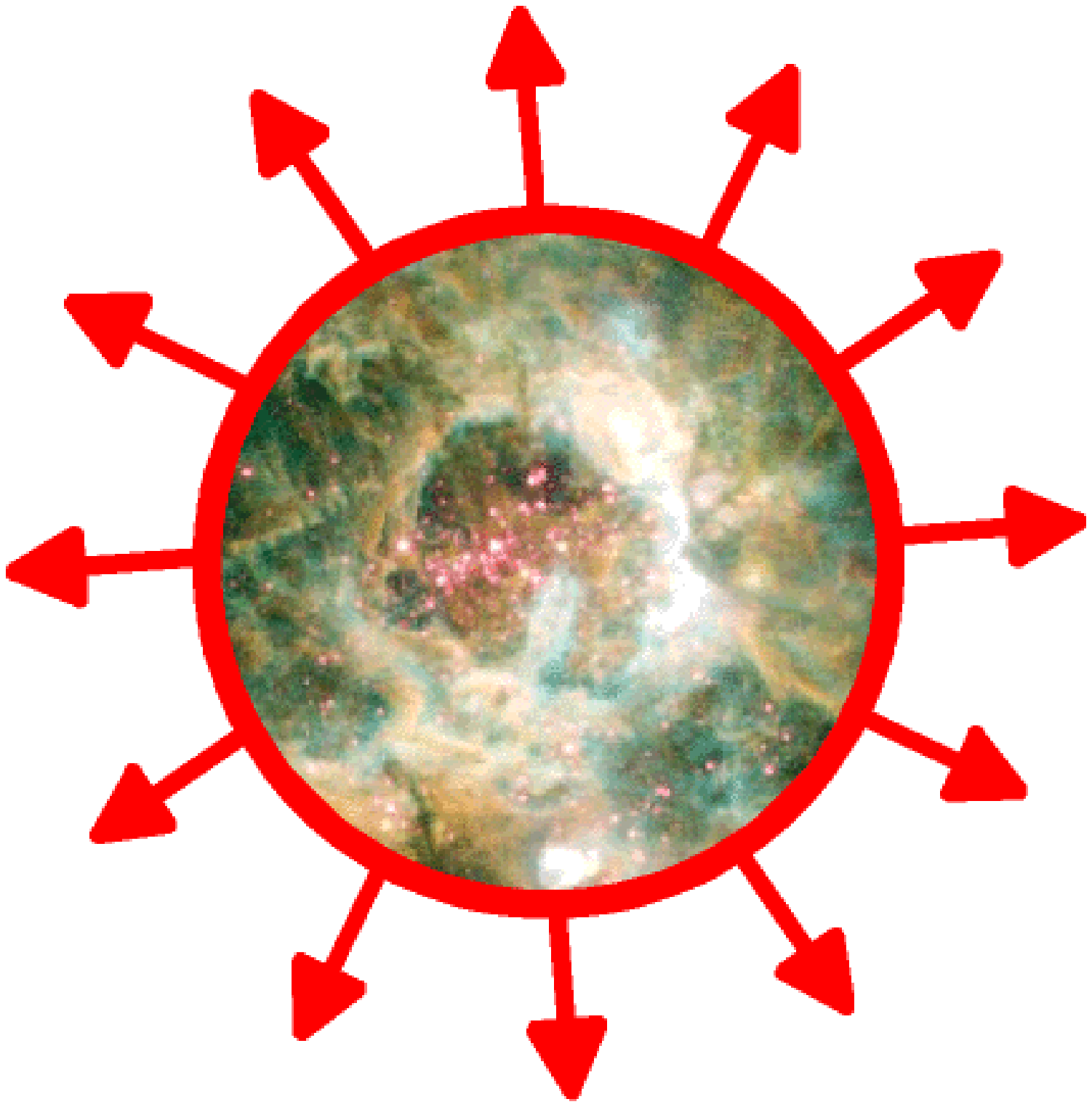,width=4.99cm,angle=-0}
 \hfill
 \psfig{file=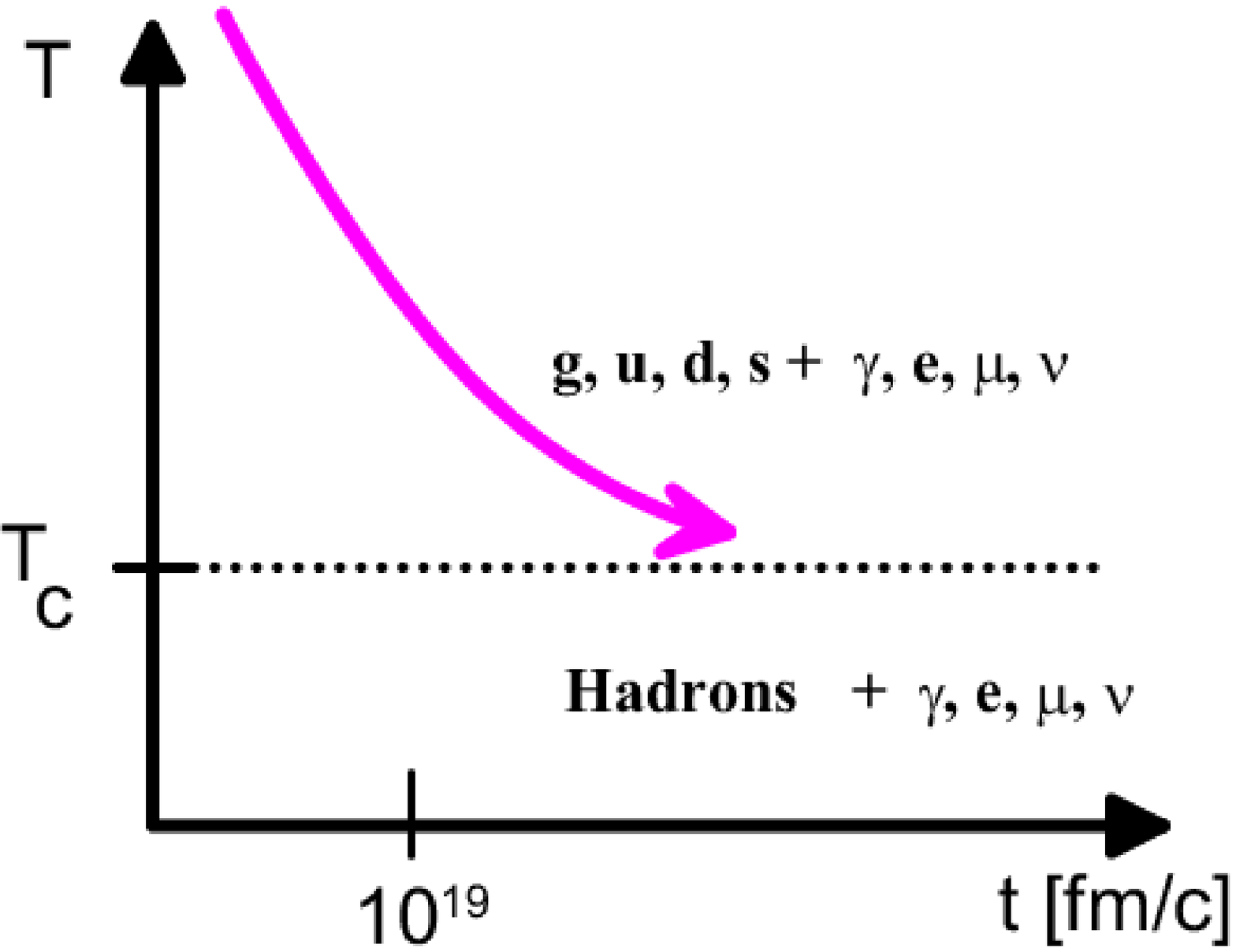,width=4.9cm,angle=-0}
 \caption{Left panel: A sketch of the expanding universe.
 Right panel: A sketch of the temperature evolution till confinement.}
 \label{Big_Bang}
\end{figure}

\subsection{Little Bang vs.\ Big Bang}

One can visualize the similarities and differences of the evolution
of matter in the Little Bang and the Big Bang as follows.
Let us consider in both
cases high enough temperatures, so that the strongly
interacting matter component is in a deconfined state.
Initially, in the Little Bang (see Fig.~\ref{Little_Bang})
the nuclear matter is in its ground state.
The uni-directional motion during the approach stage has zero entropy.
During the virulent collision this motion is randomized and a substantial
part of the kinetic beam energy is converted into particle excitations.
According to dynamical models \cite{Geiger}, after a time scale
of $\tau_1 \sim {\cal O}$(1 fm/c) or even shorter, the matter is locally
equilibrated and may be characterized by an initial or maximum temperature
$T_1$.

According to present estimates the maximum temperatures are
${\cal O}(200)$ MeV for CERN-SPS\footnote{
\underline{S}uper \underline{P}roton \underline{S}ynchrotron at CERN, i.e. the
European Laboratory for Particle Physics in Geneva}
\cite{Gallmeister},
${\cal O}(550)$ MeV for BNL-RHIC\footnote{
\underline{R}elativistic \underline{H}eavy-\underline{I}on 
\underline{C}ollider at 
\underline{B}rookhaven \underline{N}ational \underline{L}aboratory} 
and
${\cal O}(1000)$ MeV for CERN-LHC\footnote{
\underline{L}arge \underline{H}adron \underline{C}ollider at CERN} 
\cite{our_PL_1998}.
As mentioned above, already at CERN-SPS energies the body of observational
material is in agreement with the expectation of having produced a state
of strongly interacting matter where gluons and quarks are the relevant
degrees of freedom, abbreviated in the very right panel in
Fig.~\ref{Little_Bang} by $g, u, d, s$. Due to the enormous pressure
the system expands and converts into hadrons, eventually registered in
detectors. Note that, despite of the many measured particle species,
including the direct probes
\cite{direct_probes,Gallmeister_2,Gallmeister},
a simple ``yes'' or ``no'' signal, whether a deconfined state is transiently
created in heavy-ion collisions, is not at disposal. Instead, rather
a combination of various, partially subtle, observables allows the conclusion
to have produced a quark-gluon plasma. In contrast to the situation
in the cosmic evolution, under laboratory conditions the beam energy can be
changed, and within some limits the system size can be varied as also the
impact parameter.

Presumably, the evolution of matter in the Big Bang
(see Fig.~\ref{Big_Bang}) started at much
higher temperature. Note again the difference of the time scales
in Figs.~\ref{Little_Bang} and \ref{Big_Bang}. In addition, in the
Big Bang there is the non-negligible contribution of the photon and
lepton background, as indicated in Fig.~\ref{Big_Bang},
which is in kinetic and chemical equilibrium with the strongly
interacting matter.

\begin{figure}
 \vskip .01cm
 \psfig{file=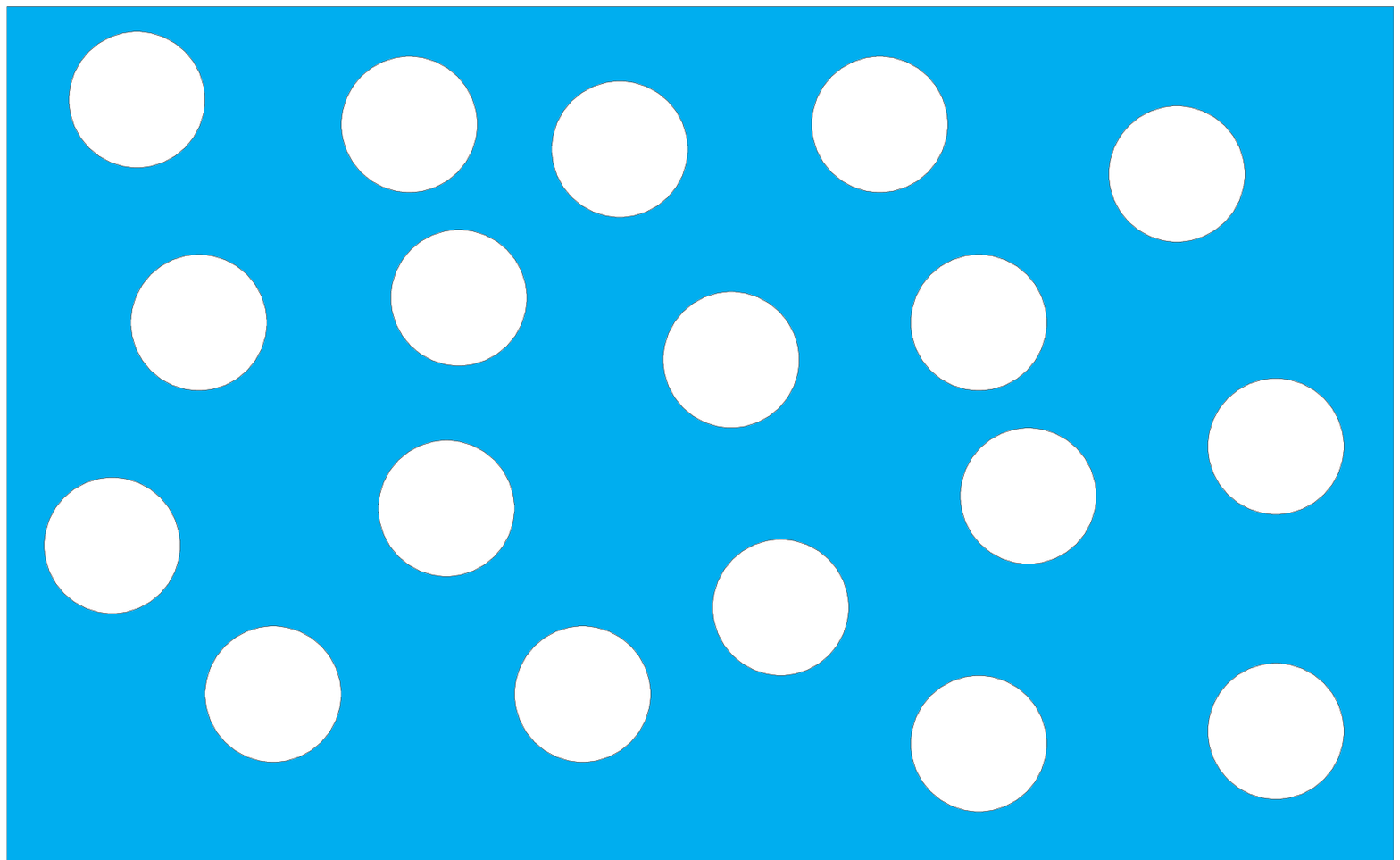,width=3.6cm,angle=-0} \hfill
 \psfig{file=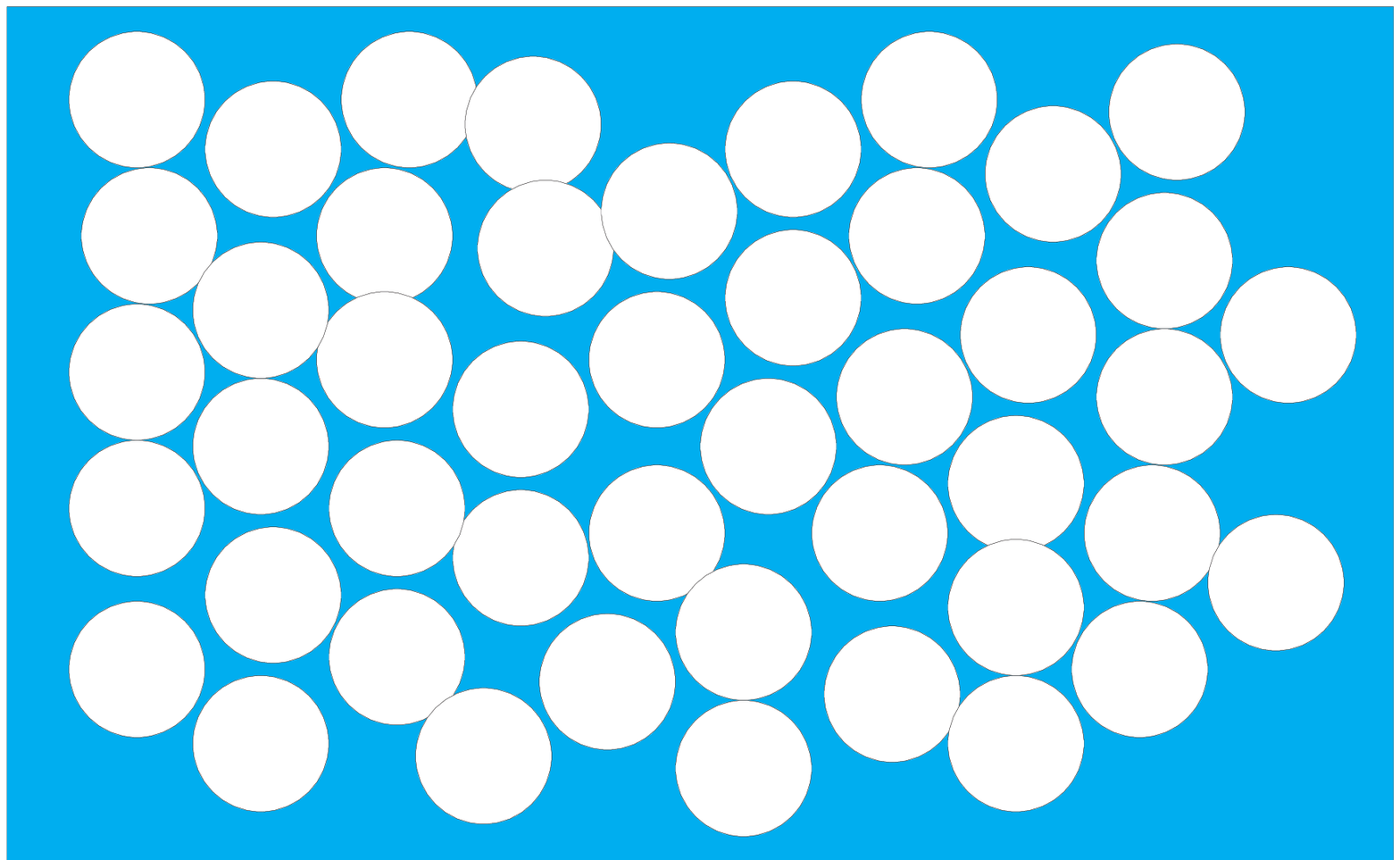,width=3.6cm,angle=-0} \hfill
 \psfig{file=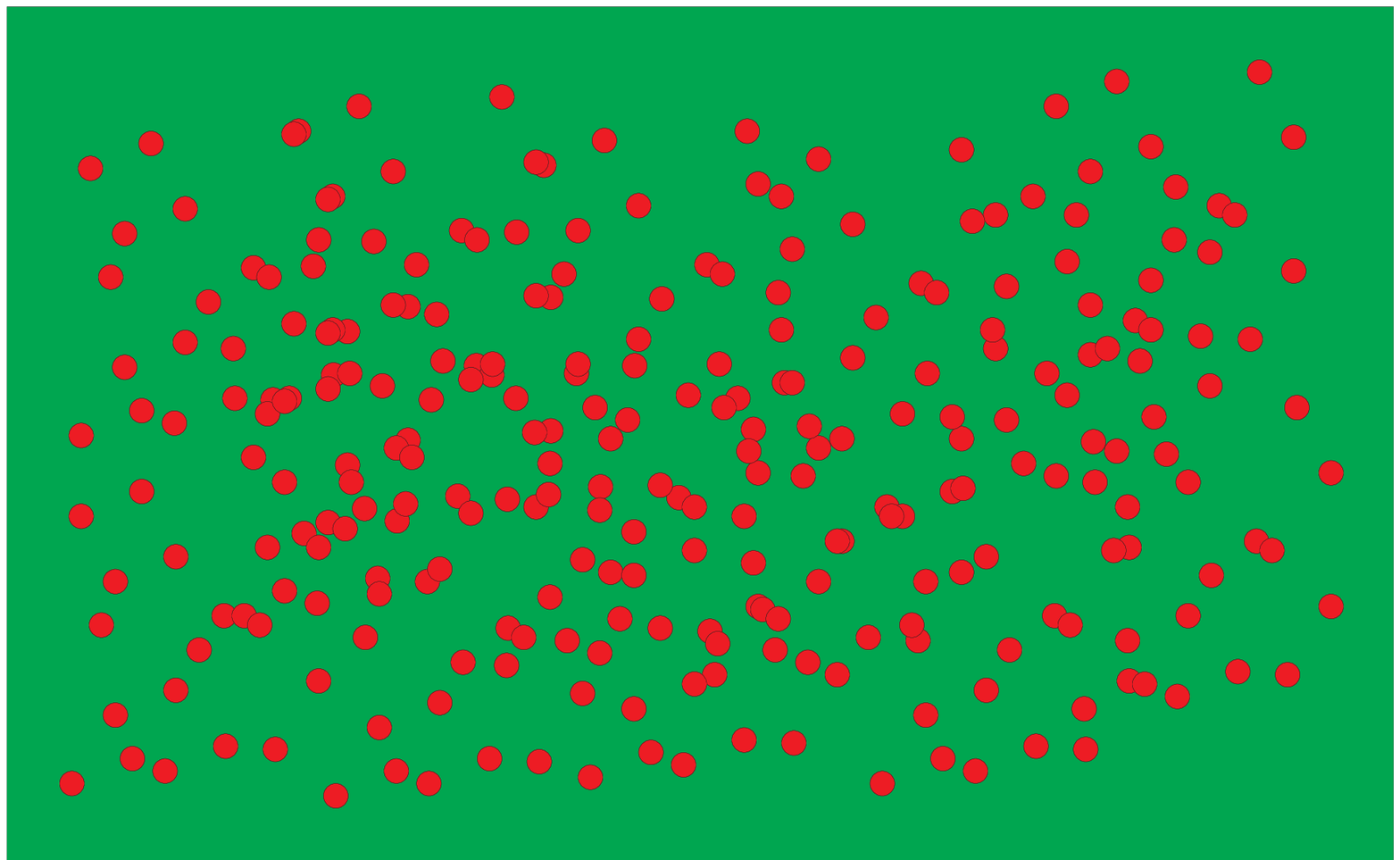,width=3.6cm,angle=-0}\\[1mm]
{} \hspace*{.8cm}  $T = 100$ MeV  \hspace*{2.3cm} $T = 150$ MeV
 \hspace*{2.3cm} $T = 200$ MeV
 \caption{A sketch of strongly interacting matter at various
 temperatures. Left and middle panels: Hadrons with finite extension
 are the relevant degrees of freedom. Right panel: Point-like quarks
 and gluons are the relevant degrees of freedom, the vacuum state
 is changed.}
 \label{dense_hadrons}
\end{figure}

There is a simple argument that dense strongly interacting matter
must change its state. Consider a hadron system at temperature of
$T \sim 100$ MeV. There are mostly pions excited with an admixture
of kaons, etas, omega and rho mesons,
and a few nucleons and their anti-particles.
At higher temperature, say around $T \sim 150$ MeV
the hadrons start touching each other
since they have a finite spatial extension
on the scale of 1 fm (see Fig.~\ref{dense_hadrons}). At slightly
higher temperatures, the hadrons cannot be longer the proper degrees
of freedom. Instead the matter consists of quarks and gluons.
Probably also the vacuum structure changes, as indicated by a different
gray scale in the background of the right panel of Fig.~\ref{dense_hadrons}.

Now, we are now interested in the dynamics of the confinement transition,
i.e., the conversion of deconfined quarks and gluons into hadrons.

\section{QCD, quarks \& gluons, confinement}

In order to describe the transition from quark-gluon matter to hadron matter
in more detail one has to rely on the proper theory of strong
interaction, i.e. QCD. It rests on the
Lagrange density (cf.\ \cite{QCD} for details, e.g.)
\begin{eqnarray}
{\cal L} = \bar \psi (i \gamma_j D^j -{\cal M}) \psi
- \frac 14 F_{ij}^a F^{ij}_a
\label{QCD_L}
\end{eqnarray}
with exact local SU(3) gauge symmetry. $D^j$ is the gauge covariant
derivative, $\psi$ and $\bar \psi$ stand for Dirac spinors
and their adjoints, $\gamma_j$ are Dirac's matrices,
and $F_{ij}^a$ denotes the field strength tensor of the
non-Abelian gauge fields. Since the different quark flavors have
different masses,
${\cal M}$ in (\ref{QCD_L}) denotes the mass matrix.

Renormalization on the one-loop level results in the running strong
coupling strength
\begin{eqnarray}
\alpha_s (Q^2) = \frac{4 \pi}{\beta_0 \ln
\left( \frac{Q^2}{\Lambda_{\rm QCD}^2} \right) }, \quad\quad
\beta_0 = 11 - \frac 23 N_f
\label{running_coupling}
\end{eqnarray}
via dimensional transmutation with the QCD scale
$\Lambda_{\rm QCD} \sim {\cal O}(200)$ MeV;
$N_f$ is the number of involved quark flavors.
In a system with a very large momentum scale $Q^2$,
the asymptotic freedom follows immediately from (\ref{running_coupling}),
i.e., $\lim_{Q^2 \to \infty} \alpha_s \to 0$.
Indeed, if in a gas of quarks and gluons the momentum scale
increases with increasing temperature, one expects asymptotically
an ideal gas. In contrast, at $T \to T_c$, the system is strongly coupled
and perturbative calculations basically fail.

\begin{figure}
 \vskip -.01cm
 \psfig{file=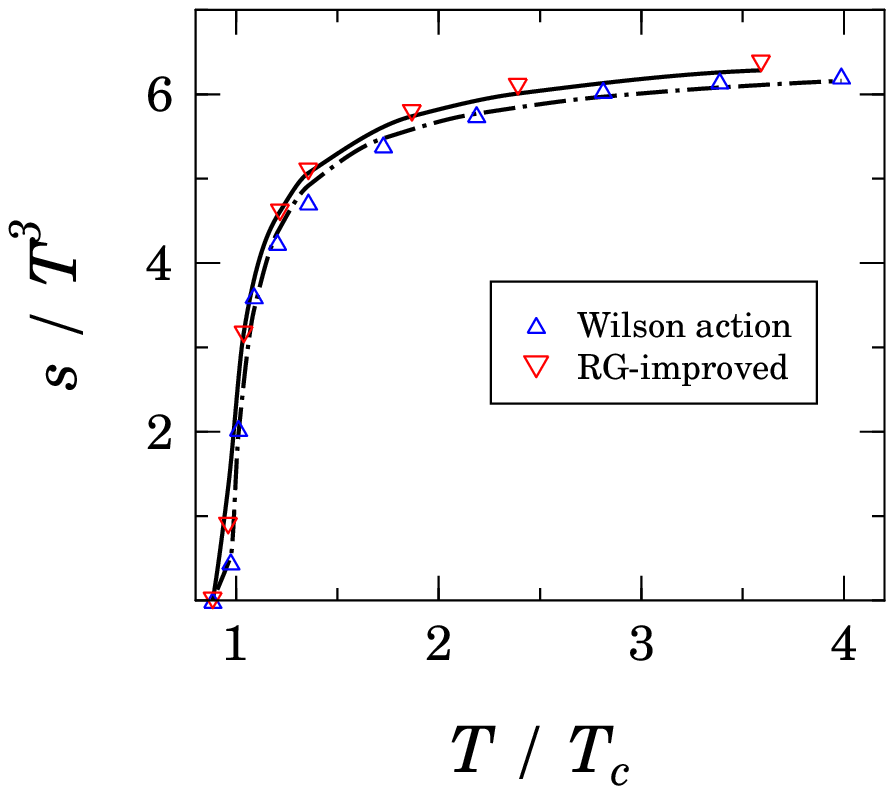,width=4.2cm,angle=-0} \hfill
 \psfig{file=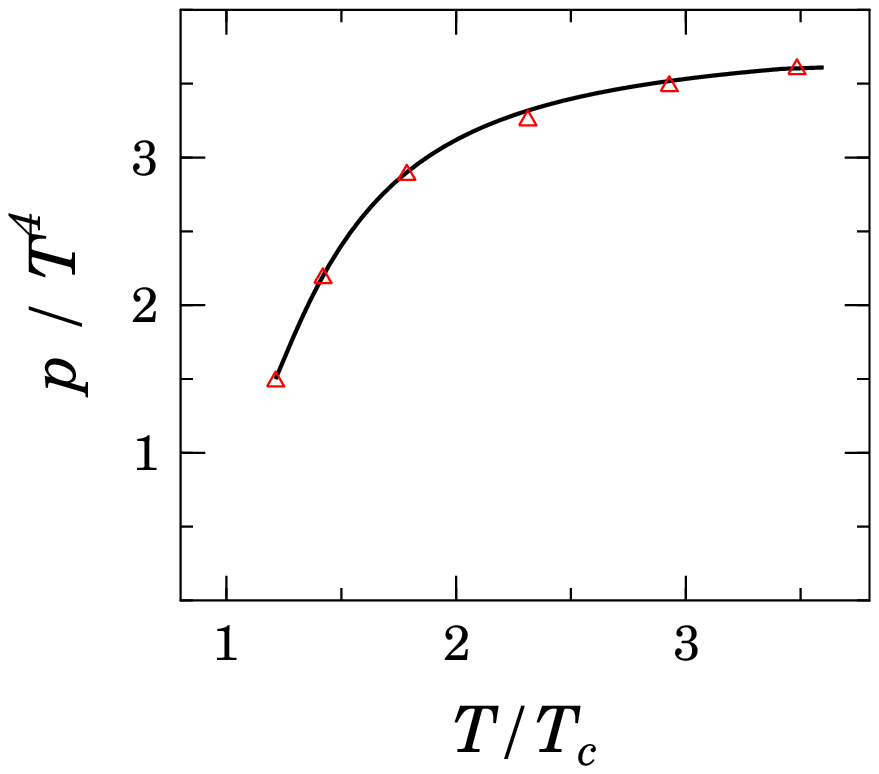,width=4.2cm,angle=-0} \hfill
 \psfig{file=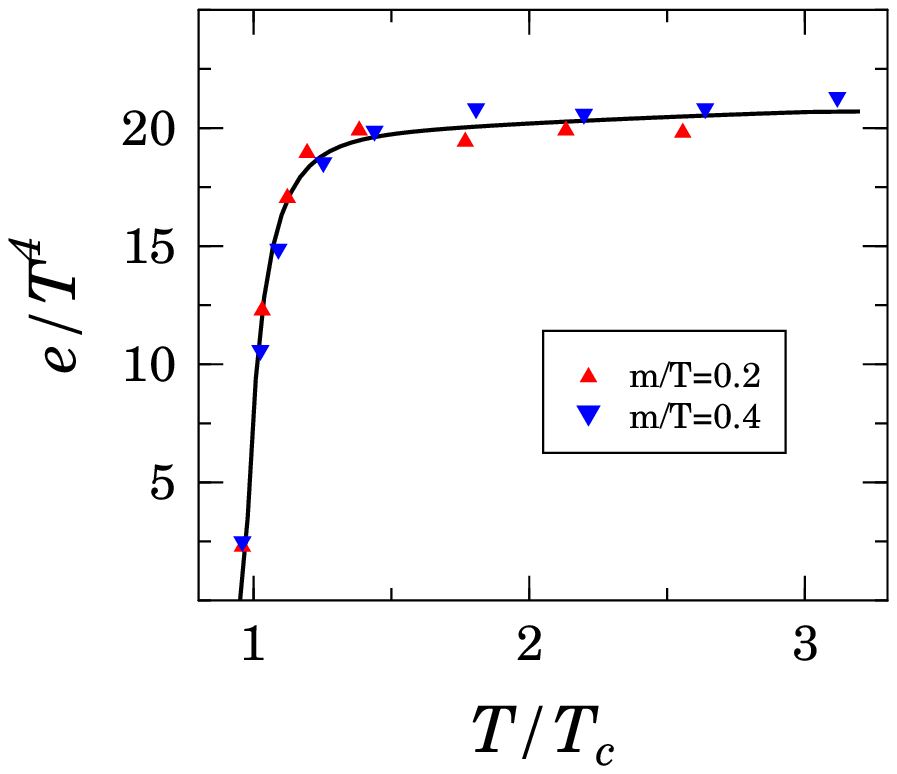,width=4.2cm,angle=-0}
 \caption{Thermodynamics of a gluon gas (left panel) and
 a two-flavor (middle panel) and a four-flavor
 (right panel) quark-gluon plasma
 as a function of the scaled temperature. Lattice QCD results
 (symbols) from the Bielefeld group. The curves represent an adjusted
 quasi-particle model, for details consult \cite{our_QCD_paper}.}
 \label{lattice_QCD}
\end{figure}

The only reliable way of {\it ab initio} calculations of thermodynamical
properties of a quark-gluon plasma is to perform Monte Carlo simulations
of finite temperature QCD in a discretized space-time.
A few results of such lattice QCD calculations are displayed in
Fig.~\ref{lattice_QCD} together with results of a quasi-particle model
\cite{our_QCD_paper} with parameters adjusted to the data.

\begin{figure}
 \vskip .01cm
 \psfig{file=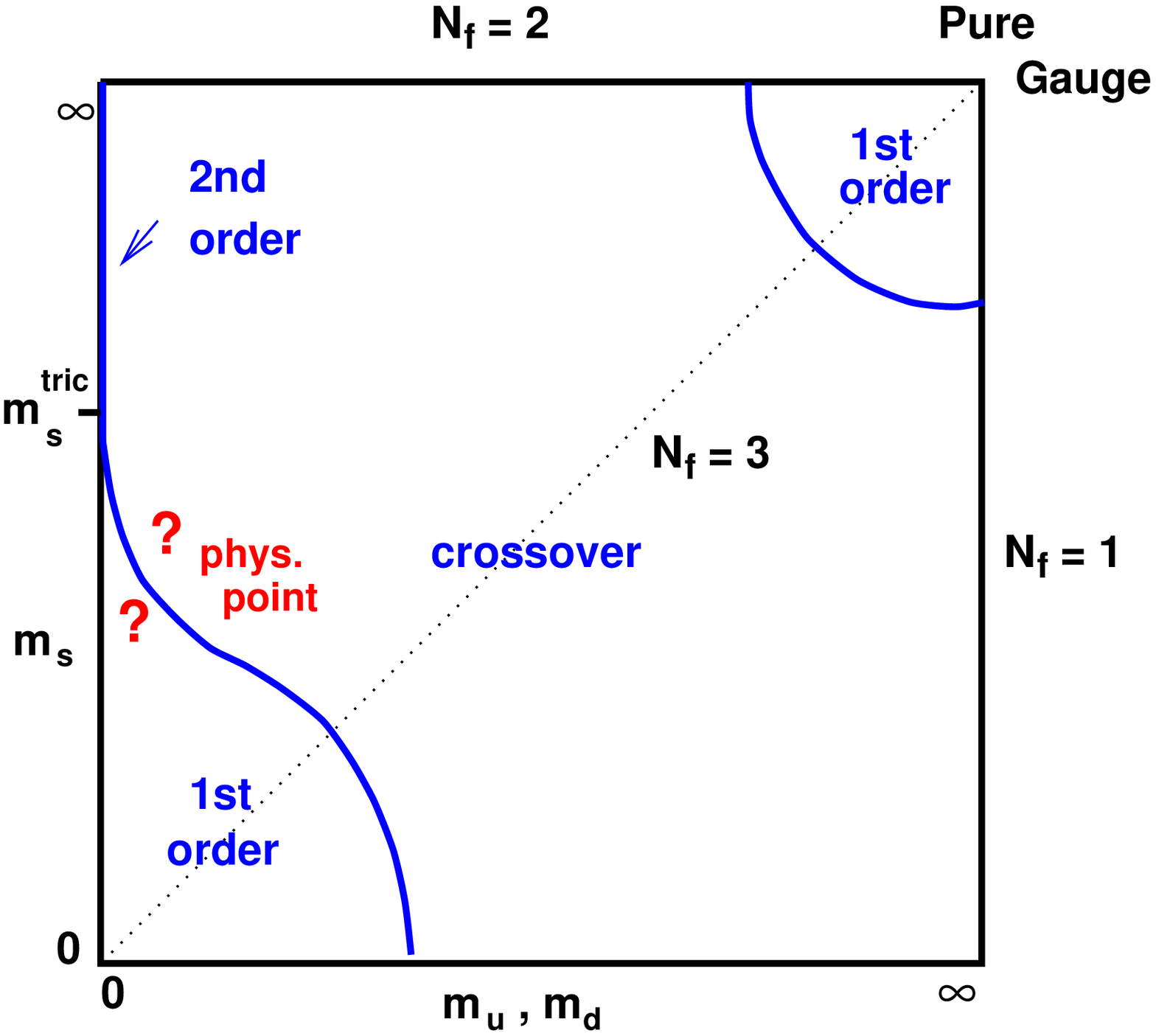,width=5.6cm,angle=-0}
 \hfill
 \psfig{file=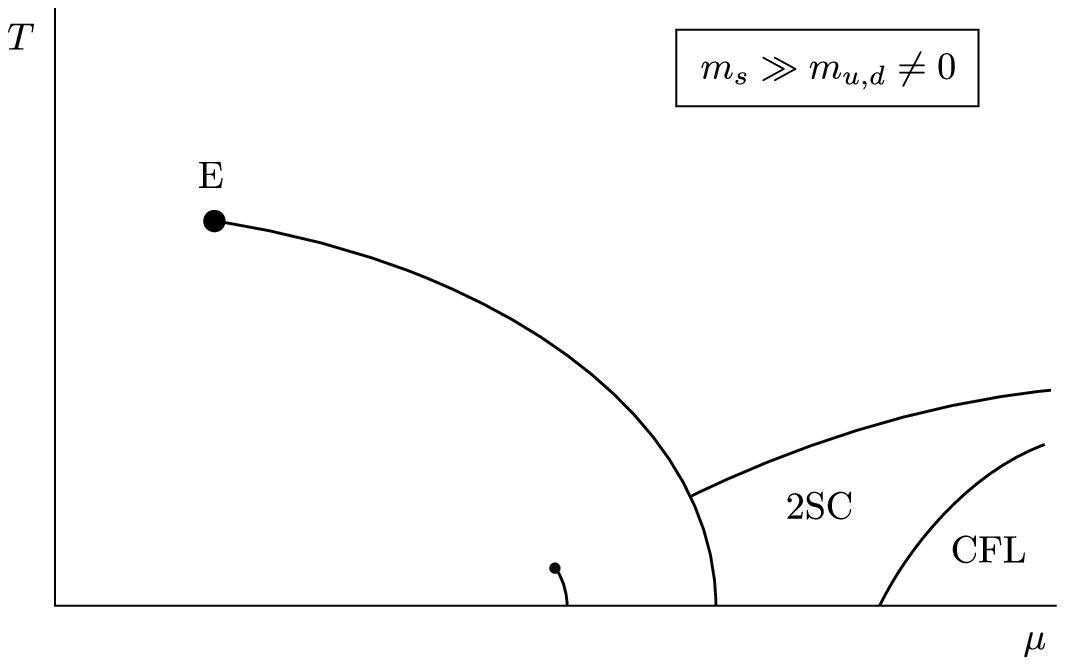,width=6.6cm,angle=-0}
 \caption{Phase diagrams of strongly interacting matter.
 Left panel: The order of the phase transition as a function of the
 quark mass parameters (from \cite{Karsch}).
 Right panel: The phase diagram in the $T - \mu$ plane according
 to \cite{Ragagopal}.
 }
 \label{order_of_transition}
\end{figure}

The transition temperature is estimated as
$T_c \sim 160$ MeV $\sim {\cal O}(m_\pi) \sim 1.6 \times 10^{12}$ K
for a quark-gluon plasma, while for a pure gluon gas it would
be  $T_c^{\rm glue} \sim 240$ MeV ($m_\pi$ is the pion mass).
Note that the present lattice QCD calculations
cover only the region $T > T_c$ and $\mu = 0$.
In order to describe the equation of state for hadron matter below $T_c$
reliably, a prerequisite would be
an accurate description of the mass spectrum of the hadrons, a goal
not yet fully accomplished. Despite the lack of an equation of state
at $T < T_c$, some information on the order of the transition is available.
Inspection of the left panel of
Fig.~\ref{order_of_transition} reveals that if the masses of the light
current quarks are sufficiently small
the deconfinement transition is of first order.
The physical relevant case of the two light flavors $u, d$ and a
medium-heavy $s$ quark is near to the borderline to a cross over.
The latter notion means a dramatic, but nevertheless smooth
change of thermodynamic properties in a narrow region around $T_c$.
At sufficiently small $u,d$ quark masses
and large $s$ quark mass, the phase transition would be
of second order; the regions of first order and second order are
separated by a tri-critical point. The pure gauge sector (i.e.,
infinitely heavy quarks so that they do not longer act as dynamical objects)
is known to display a first-order phase transition again.

\begin{figure}
 \vskip -.01cm
 \center
 \epsfxsize=.5 \hsize \epsffile{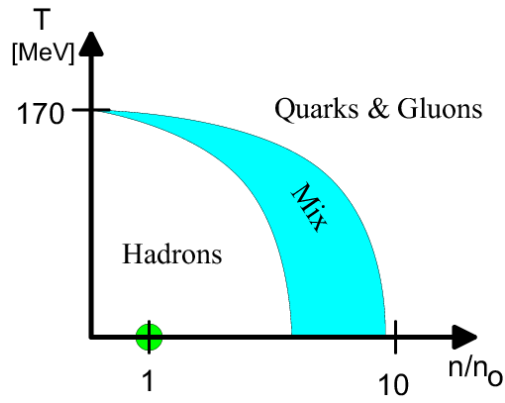}
 \caption{Naive phase diagram of strongly interacting matter
 in the $T - n$ plane ($n$ is the baryon density).}
 \label{phase_diagram}
\end{figure}

Since in the early universe the baryo-chemical potential is small
the information on the region $\mu > 0$ is not important. Otherwise,
in the Little Bang (or more importantly, in massive neutron stars with
quark cores) the finite baryo-chemical potential $\mu$ becomes important.
As seen
in the right panel of Fig.~\ref{order_of_transition}, a richer structure
of the phase diagram is showing up with di-quark condensates (2SC)
and color-flavor locking effects (CFL); also an endpoint (E) of the
phase borderline may appear at finite values of $T$ and $\mu$
(for more details, see \cite{Ragagopal}).
The latter peculiarities make the
previous naive phase diagram, as shown in Fig.~\ref{phase_diagram},
more involved.

To quote some physical numbers characterizing the nuclear matter 
near ground state one can refer to
$n_0 = 0.15 \, \mbox{fm}^{-3} = 1.5 \times 10^{38} \, \mbox{cm}^{-3}$,
$e_0 = 2.7 \times 10^{14} \, \mbox{g} \, \mbox{cm}^{-3}$,
1 MeV $\approx 10^{10}$ K.

The deconfinement transition is intimately related to chiral symmetry
restoration at temperature $T_c^\chi$.
In hadronic matter, at $T < T_c^\chi$, the effective
degrees of freedom are massive quarks and we have finite expectation
values of condensates, like the chiral condensate
$\langle \bar \psi \psi \rangle$. For
$T > T_c^\chi$ the quarks become massless and the condensates disappear.
For QCD it turns out that deconfinement and chiral symmetry restoration
coincide, i.e. $T_c = T_c^\chi$. Therefore, the chiral condensate
may serve as order parameter describing the confinement.

\section{A constructed phase transition of first order}

\begin{figure}
 \vskip .01cm
 \psfig{file=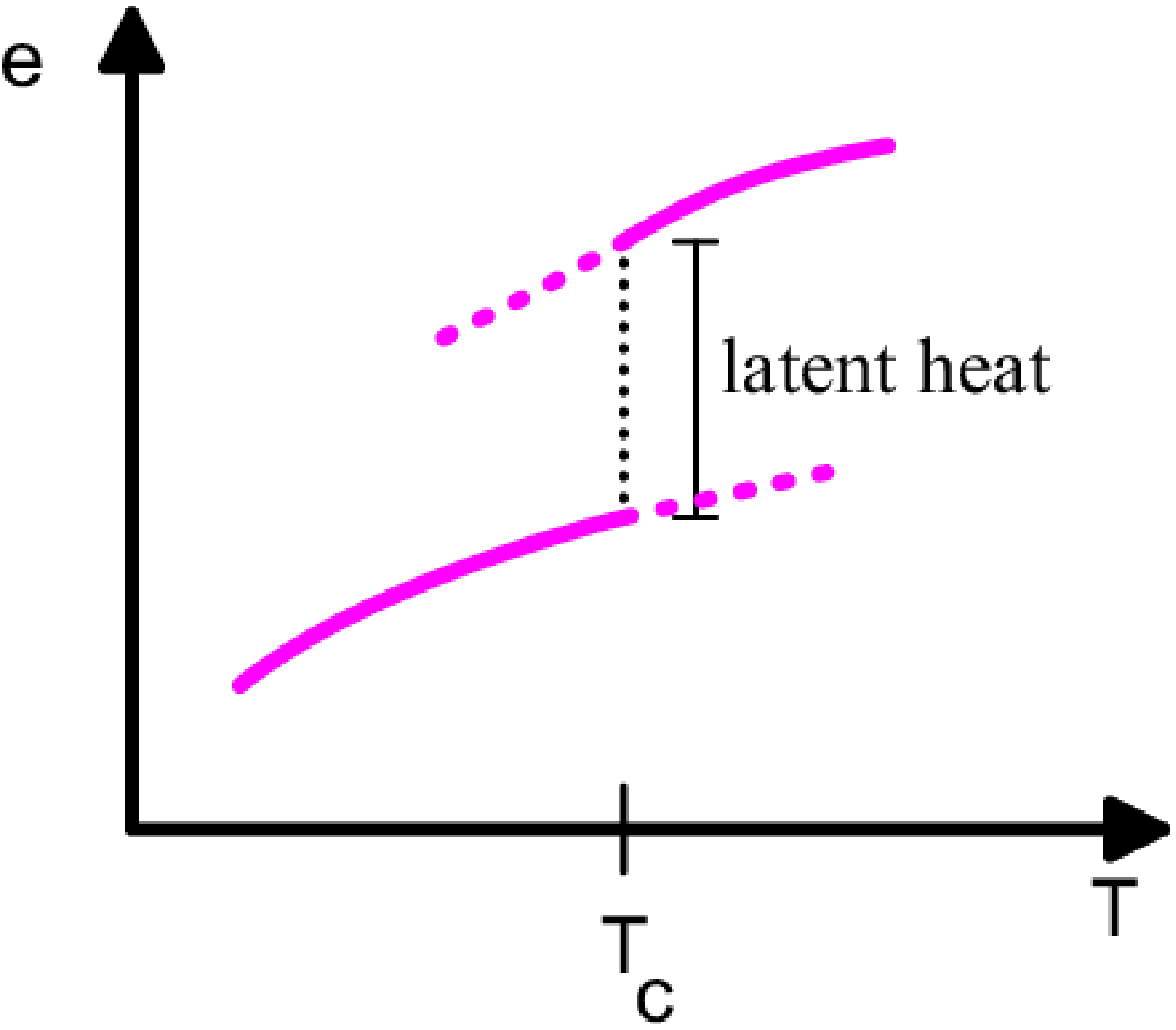,width=4.6cm,angle=-0} \hfill
 \psfig{file=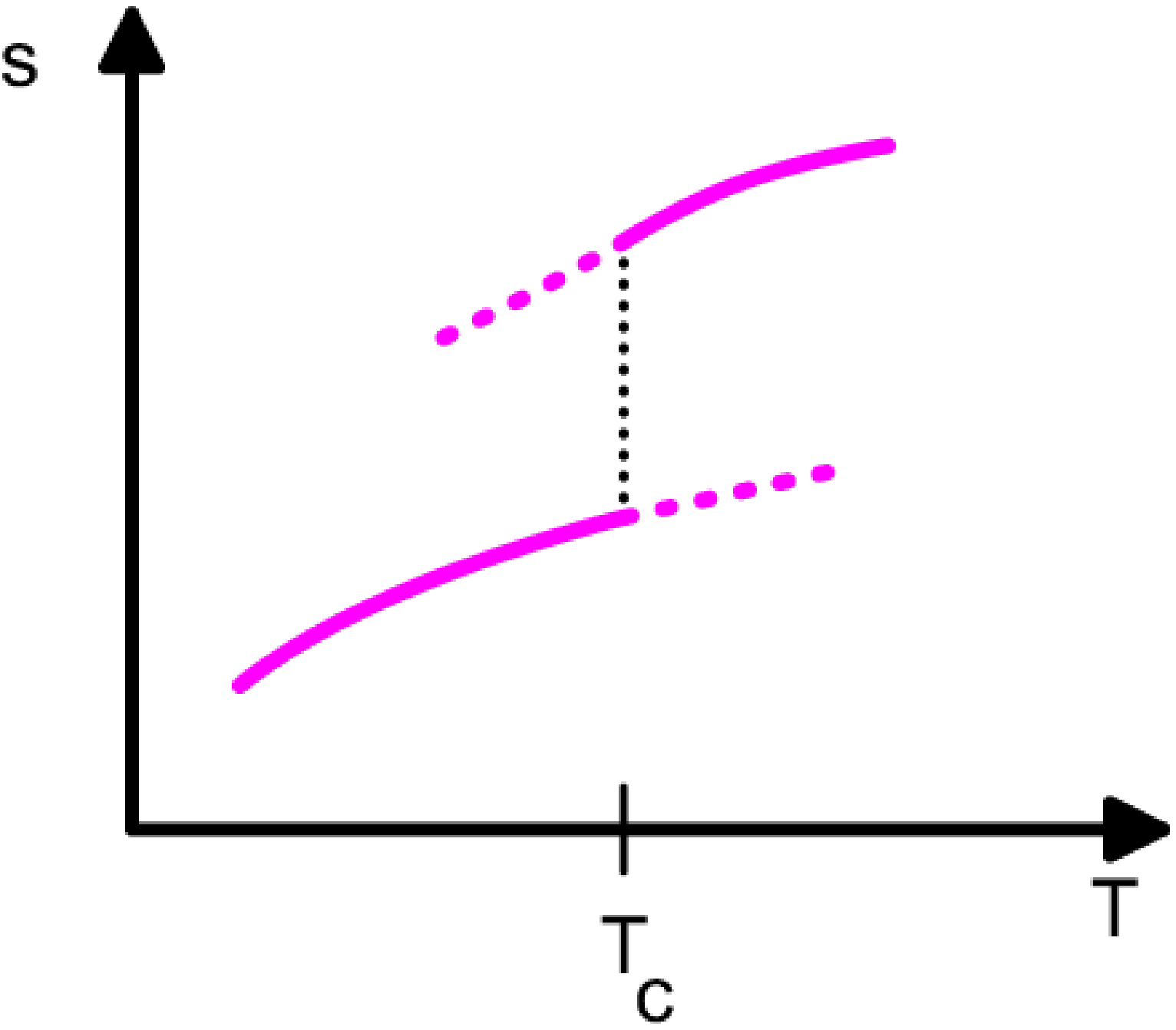,width=4.6cm,angle=-0}
 \vfill
 \psfig{file=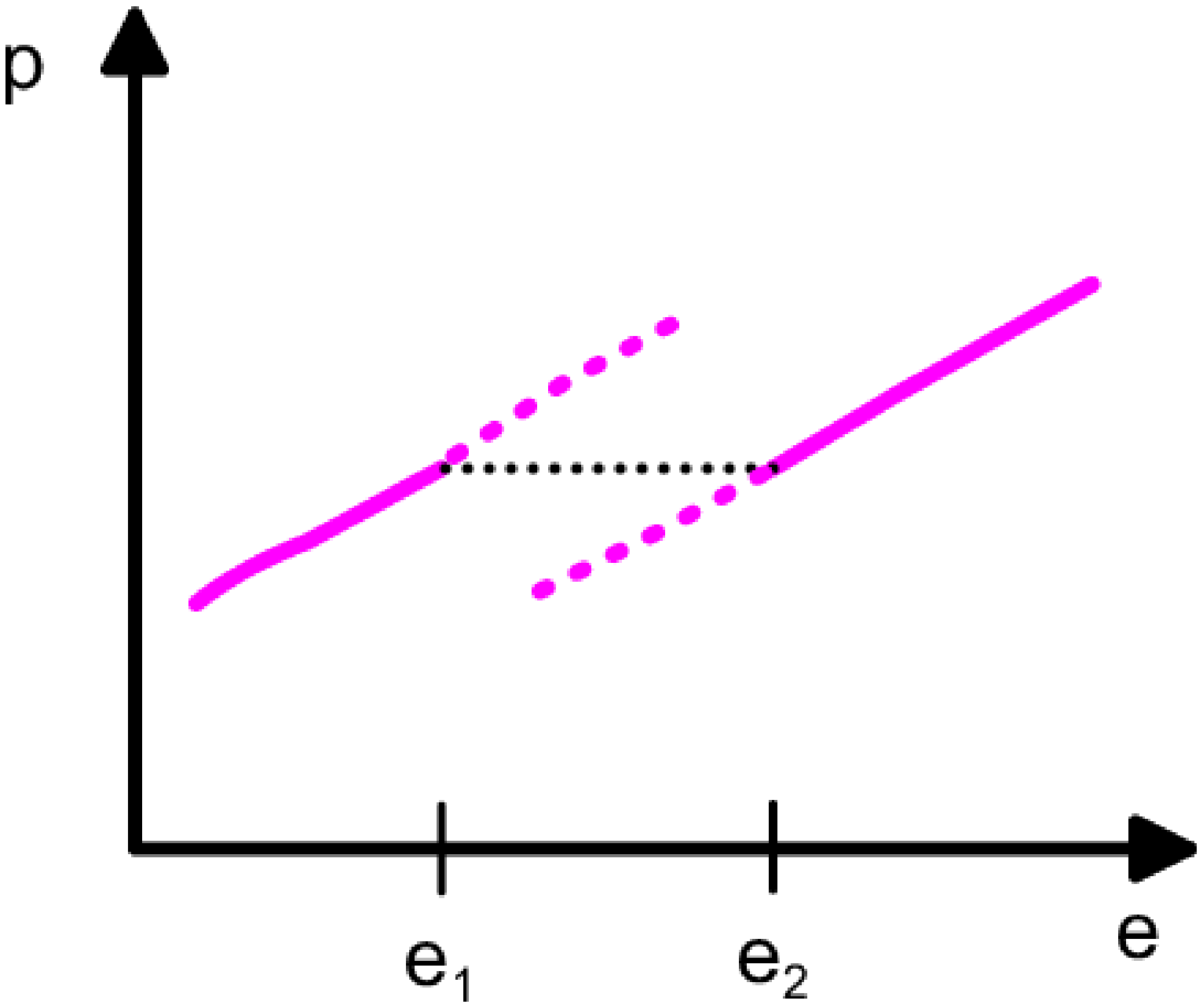,width=4.6cm,angle=-0} \hfill
 \psfig{file=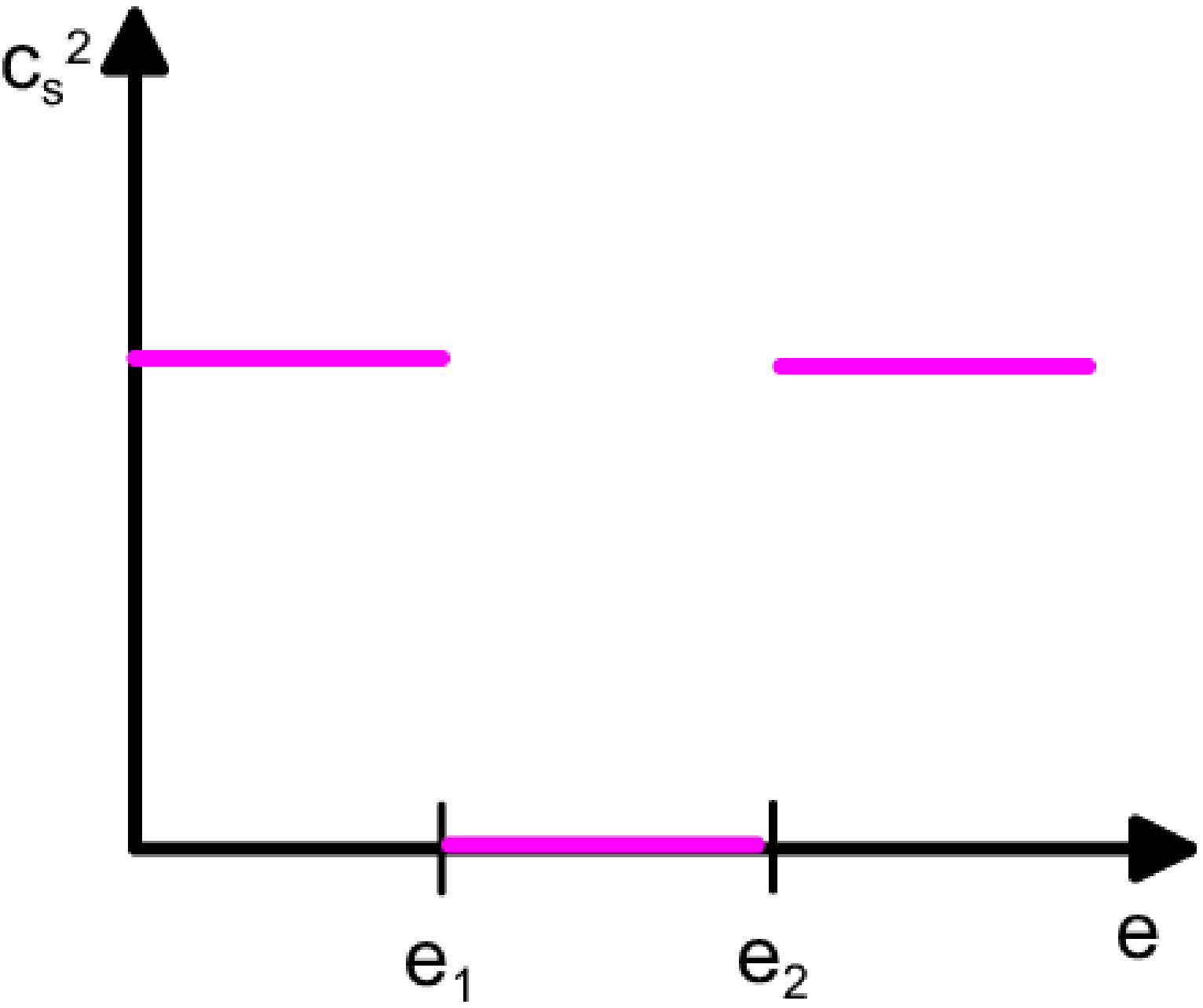,width=4.6cm,angle=-0}
 \caption{Schematic sketch of the thermodynamic state variables for a
 strong first-order phase transition. Upper row:
 Energy density $e$ and entropy
 density $s$ as a function of the temperature.
 Lower row: Pressure $p$ and sound velocity
 $c_s^2 = \partial p / \partial e$ as a function of the energy density.}
 \label{eos_thermodynamics}
\end{figure}

In view of the above discussion of the order of the deconfinement
transition, it seems legitimate to approximate the thermodynamics in
the transition region by a suitable construction.
This construction results in a strong first-order phase transition
with large latent heat (cf.\ Fig~\ref{eos_thermodynamics})
which probably overestimates of what we can expect in reality.
The construction, however, is made quite general so that it might be
applied also in other cosmic phase transitions, such as on
electroweak and GUT scales to be discussed below.
A MIT-type bag model is quite popular:
asymptotically free quarks are confined in a certain
spatial region, called bag. The vacuum energy density in the bag is enlarged
by a constant amount of $B$ when compared with ordinary vacuum.
The equation of state reads then
\begin{eqnarray}
p_{\rm QCD} = d_{g,u,d} \frac{\pi^2}{90} T^4 - B,
\quad\quad d_{g,u,d} = 37,
\label{bag_eos}
\end{eqnarray}
with vacuum pressure $-B$. The use of the pure radiation term
$\propto T^4$ is suggested by asymptotic freedom; near to $T_c$,
the non-perturbative effects are thought to be parameterized by $B$.
This ansatz allows only a poor reproduction of the lattice
QCD results, and for more detailed considerations it must be replaced
by other models like that in \cite{our_QCD_paper}.
In some sense this schematic equation of state can be considered
as a down-extrapolation from asymptotic freedom.
Similarly, one can up-extrapolate the hadron equation of state,
and the most simple ansatz is that of a pion gas
neglecting the strong interaction:
\begin{eqnarray}
p_\pi = d_\pi \frac{\pi^2}{90} T^4, \quad\quad d_\pi = 3.
\label{pion_eos}
\end{eqnarray}
Also this too simple model needs improvements by including the other
hadrons (cf.\ \cite{hadron_eos_1}) and their strong interactions.
In \cite{hadron_eos_2} the interested reader can find a selection
of some recent approaches and various methods. Despite of many efforts,
a generally accepted nuclear or hadronic equation of state at large
temperature and density has not yet emerged. A direct comparison of the various
model equations of state with experimental data from heavy-ion collisions
at BEVALAC, SIS and AGS energies (i.e., beam energies of $1 \cdots 2$
and $10 \cdots 15$ A$\cdot$GeV)
is hampered by dynamical and non-equilibrium
effects \cite{Stoecker_Cassing}. A recent survey can be found in
\cite{eos2000}.

\begin{figure}
 \vskip .01cm
 \center
 \psfig{file=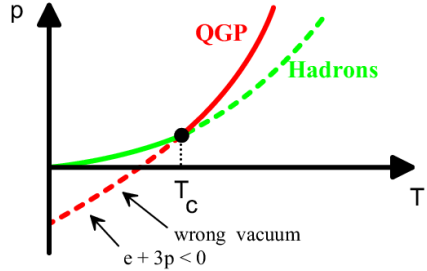,width=5.5cm,angle=-0}
 \caption{A constructed phase transition of first order by extrapolating
 asymptotic equations of states. The dashed branches are metastable.}
 \label{constructed_phase_transition}
\end{figure}

The equations of state (\ref{bag_eos}, \ref{pion_eos}) are displayed in
Fig.~\ref{constructed_phase_transition}. At the crossing of both curves, the
Gibbs conditions of phase equilibrium are fulfilled, i.e.,
\begin{eqnarray}
p_1 & = & p_2,\\
T_1 & = & T_2,\\
\mu_1^{(1)} & = & \mu_2^{(1)},\\
\mu_1^{(2)} & = & \mu_2^{(2)},\\
 & \cdots & .\nonumber
\end{eqnarray}
Here we consider the special case of $\mu^{(\alpha)} = 0$.
In the case of competing phases, the stable one is that
with lower free energy, which is the negative of pressure.
The dashed
sections in Fig.~\ref{constructed_phase_transition} are, therefore,
metastable branches.
According to the Gibbs criteria,
the conditions for phase equilibrium are fulfilled
at $T_c = \frac{90}{34 \pi^2} B^{1/4}$.

Fig.~\ref{eos_thermodynamics} sketches the other thermodynamical
state variables as a function of the temperature and the energy density.

Having the two asymptotic branches (\ref{bag_eos}, \ref{pion_eos}) at hand,
one can find, of course, also smooth interpolations between them
\cite{Blaizot_Ollitraut} displaying a crossover. Or, as exercised in
\cite{our_booklet}, one can construct another interpolation with a loop
structure in the potential $p(T)$; this yields two metastable and one
unstable branches, as known from the famous Maxwell or double-tangent
constructions. Thereby the strong phase transition
of first order constructed above becomes weaker.
However, the mentioned procedures are
{\it ad hoc} and need to be checked carefully against the results of lattice
QCD calculations.

\section{Phase transformation dynamics\label{nucleation_theory}}

In spite of the simplified nature of the phase transition constructed
in the previous section, we study its implications for the cosmic
evolution. As stressed above, this construction overestimates the strength
of the transition. Therefore, this model serves as an upper bound
of possible interesting effects caused by the confinement transition.

Our discussion is based on the classical nucleation theory, where
the central quantity is the critical bubble of the newly forming phase.
The change of the Helmholtz free energy can be expanded into a power
series with respect to the bubble radius $r$ as
\begin{eqnarray}
\Delta F(r,T) = - \frac{4 \pi}{3} \Delta p \, r^3 + 4 \pi \sigma r^2 -
8 \pi \gamma r + \cdots
\end{eqnarray}
with $\Delta p = p_1 - p_2$ as pressure
difference of both phases, $\sigma$ is the surface tension,
and $\gamma$ the curvature parameter. Neglecting the latter one,
the resulting critical radius,
related to the maximum of the free Helmholtz energy, is
$R_c = \frac{2 \sigma}{p_1 - p_2}$; smaller bubbles are energetically
unfavorable since the surface energy is too large in comparison with the
volume energy, while larger bubbles are too rare to influence the
phase transformation dynamics. Critical bubbles are the ones which,
after being created by fluctuations, can grow further.
Notice that the surface tension $\sigma$ is the decisive quantity governing
the transition dynamics.

The probability for forming a critical bubble
per time and volume unit is given by
\begin{eqnarray}
w = w_0 \exp\left\{
\frac{- 16 \pi \sigma^3}{ 3 T (p_1 - p_2 )^2} \right\},
\label{w_0}
\end{eqnarray}
where, from dimensional reasons,
\begin{eqnarray}
\sigma = \sigma_0 T_c^3,
\quad\quad
w_0 = \bar w_0 T_c^4
\label{sigma_0}
\end{eqnarray}
must hold, since the relevant scale is given by $T_c$.
One usually assumes that
$\sigma_0, \bar w_0 \sim {\cal O}(1)$. The calculation of the
coefficient $\bar w_0$ is often debated (cf.\ \cite{Csernai_Kapusta}), but
due to the enormously sensitive temperature dependence of the exponent
in (\ref{w_0}), changes in $\bar w_0$ may be absorbed in a redefinition
of $\sigma_0$ to be discussed below.

Once a critical bubble is created, it grows with velocity
$v = v(T, \cdots)$, where the dots indicate a possible dependence
on other state variables or transport coefficients.

Now, we derive an equation for the occupancy of the space with the new
phase. Thereby, we must take into account that bubbles created at an earlier
instant have grown and that later a reduced volume for bubble
nucleation is available. The general growth law
$x = x(t, w, v, R_c, \cdots )$ has therefore also memory effects.
Several approximations are discussed in \cite{our_booklet}.
Here we take the so-called Avrami approximation
\begin{eqnarray}
x_2 = 1 - \exp\{-h\},
\quad\quad
h = \frac{4 \pi}{3} \int_0^t d \tau \, w(\tau) R(\tau)^3
\left[\frac{R_c(\tau)}{ R(\tau)}
+ \int_\tau^t d \theta \, \frac{v(\theta)}{R(\theta)}\right]^3,
\label{Avrami}
\end{eqnarray}
where $x_2$ is the volume weight of phase 2.

\begin{figure}
 \vskip .01cm
 \centerline{ \psfig{file=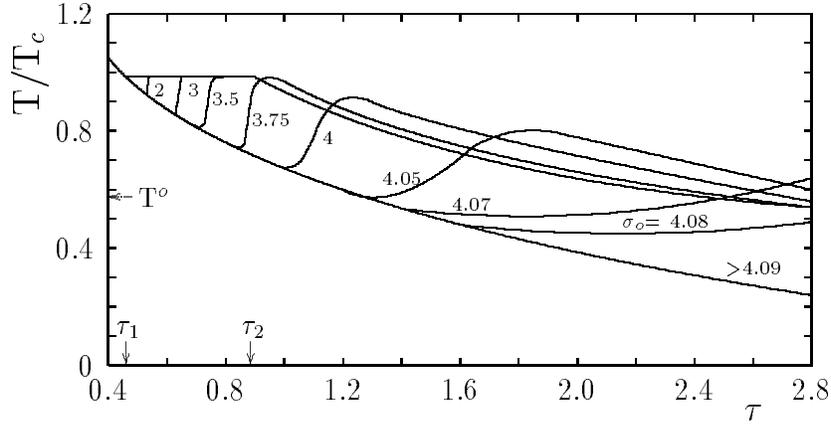,width=11.1cm,angle=-0}}
 \caption{The temperature evolution during a first-order phase transition
 for various values of the surface tension parameter $\sigma_0$
 as a function of the scaled dimensionless time
 $\tau = 2 {\cal C} B^{1/2} t$.
 $\tau_1$ and $\tau_2$ denote the beginning and end of an equilibrium
 transition; $T^o$ is the temperature of the maximum nucleation rate
 (\protect\ref{w_0}).}
 \label{T(t)}
\end{figure}

Fig.~\ref{T(t)} displays the outcome of a numerical study of the temperature
evolution during confinement. The equations of state
(\ref{bag_eos}, \ref{pion_eos}) are supplemented by the background contribution
$p_{\rm bg} = d_{\rm bg} \frac{\pi^2}{90} T^4$, with $d_{\rm bg} = 14.25$.
One observes for vanishing surface tension a quasi-equilibrium transition
with negligible supercooling, see the flat section in Fig.~\ref{T(t)}.
Increasing the surface tension parameter
$\sigma_0$ causes a deeper supercooling. Then massive nucleation of
many critical bubbles happens. This causes a sudden reheating to $T_c$,
due to the released latent heat. Then nucleation ceases and the transition
proceeds via bubble growth. Further increasing the surface tension smoothes
out the transition and leads to a longer duration of the transition era.
In the extreme case, hypercooling happens, where the system stays below
$T_c$.

To summarize the typical scales of the
confinement transition in the Big Bang let us mention that
the estimates of the
transition temperature $T_c \sim 160$ MeV from lattice QCD yield:
\begin{itemize}
\item
beginning of the transition at world age $t_1 \sim 6$ $\mu$sec,
\item
end of the transition in case of a near to equilibrium transition
with small surface tension
$t_2 \sim 12$ $\mu$sec.
\end{itemize}
Other characteristic quantities are
\begin{itemize}
\item
horizon radius $R_H \sim 10$ km,
\item
Hubble time $t_H \sim 10^{-5}$ sec,
\item
energy density within the horizon $M_H corresponding to \sim 1 M_\odot$,
\item
baryon charge within horizon $N_H^B \sim 10^{50}$,
\item
$e_{CDM} \sim 10^{-8} e_{rad}$
(from a back-extrapolation of estimates of the present
cold dark matter energy density).
\end{itemize}

\section{Details of the cosmic confinement transition}

\subsection{Small supercooling \label{small_supercooling}}

Early QCD lattice calculations predicted an astonishingly small
surface tension (\ref{sigma_0})
between hadronic and deconfined matter: $\sigma_0 =$
0.0292 \cite{surface_tension_1},
0.0155 \cite{surface_tension_2},
0.014 $\cdots$ 0.03 (or 0.44) \cite{surface_tension_3}.
As these numbers are valid for pure SU(3) gauge theory, where the deconfinement
transition is undoubtedly of first order, one can speculate whether this
feature of $\sigma_0 \ll 1$ is pertinent when
the quark degrees of freedom are included. As a consequence, the transition
is expected to proceed with small initial supercooling,
$\Delta T / T_c \le 10^{-3}$, and follows then the unlabeled flat section
in Fig.~\ref{T(t)}.
The temperature is kept constant at $T_c$ due to the
condensation heat released. Therefore, the transition resembles an
equilibrium transition since bubbles have time enough to grow and
the released latent heat is rapidly distributed uniformly.
Entropy production is negligible.
The expansion of the scale factor is slightly changed
and points to a mini-inflationary era \cite{mini_inflation},
i.e., an expansion faster than $R \propto \sqrt{t}$.

An opposite scenario is claimed in \cite{Cottingham}. There, within
an effective two-phase model with strong first-order phase transition,
the possibility of a substantial supercooling is considered and, as a
consequence, an exponential growth of the scale factor and a huge entropy
increase are advocated. The entropy increase dilutes the baryon-to-photon
ratio. For instance, a pre-fabricated ratio $n_B/n_\gamma \sim {\cal O}(1)$
prior to the transition can be diluted to the needed value of $10^{-9}$
(prior to nucleosynthesis) if the entropy increase is tuned to
$10^9$. Scenarios of such a type have been elaborated for
inflation and will be discussed in section \ref{Inflation}.
We consider Fig.~\ref{order_of_transition} and notice the small values of the
surface tension from lattice QCD.
Then the scenario in \cite{Cottingham} seems to be too extreme.
In particular, during the rapid expansion of matter in the Little Bang
such parameters would cause an extremely deep supercooling, in contrast
to the expectation \cite{rapid_nucleation}.

To conclude this subsection, we mention a series of papers \cite{Toki},
in which the confinement transition is studied within an effective theory,
namely the dual Ginzburg-Landau model based on the dual Higgs mechanism
by QCD monopole condensation.

\subsection{Turbulent confinement transition?}

\begin{figure}
 \vskip -.01cm
 \center
 \psfig{file=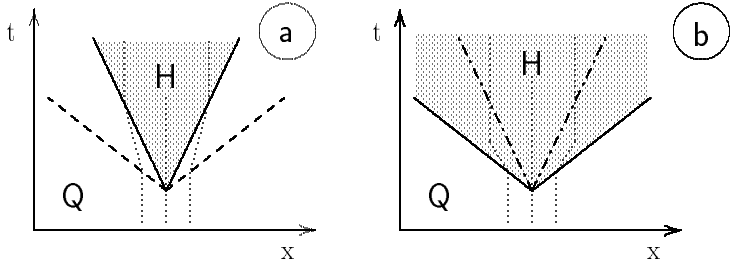,width=9.9cm,angle=-0}
 \caption{
 Space-time diagram of
 two types of growing bubbles of hadronic matter (H) in deconfined matter (Q).
 Left panel: Deflagration bubbles.
 Right panel: Detonation bubbles.
 The solid lines are the phase boundaries, while the dashed lines depict
 shock waves reheating the matter.
 The dotted lines indicate stream lines of matter.
 These sketches are strongly stretched in $x$-direction so that the
 light cone is below the dashed lines in (a) and below the solid lines
 in (b).
 }
 \label{growing_bubbles}
\end{figure}

\begin{figure}
 \vskip -.01cm
 \center
 \psfig{file=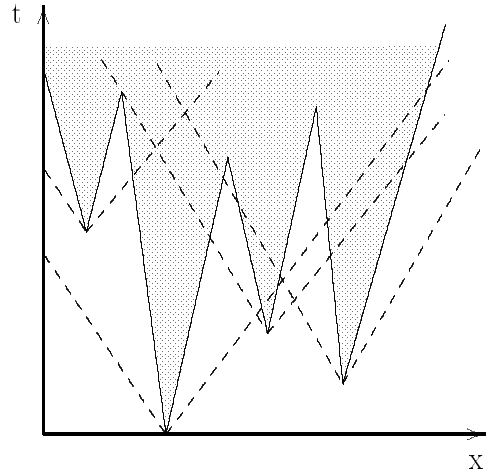,width=7.1cm,angle=-0}
 \caption{Schematic view on the process of filling the space with the confined
 phase by bubble growth via deflagration. Many intersections
 of shock waves arise.}
 \label{shocks}
\end{figure}

There are two types of hydrodynamical solutions for growing bubbles
in matter being at rest far enough from the bubble \cite{bubble_growth}, as
depicted in Fig.~\ref{growing_bubbles}: Deflagration bubbles and
detonation bubbles. In both types,
the bubble growth\footnote{See also \cite{Rezzolla} for the bubble
dynamics at the end of the cosmological confinement transition.}
is accompanied by shock waves.
If we have deflagration bubbles, these waves propagate in the deconfined
phase. Otherwise, the waves occur in the hadronic phase.
As suggested by Fig.~\ref{shocks},
there are many intersections of the shock waves during the completion
of the confinement transition. The intersections are related to
density inhomogeneities which we will discuss below.
The density inhomogeneities can represent nuclei for black hole formation
visible as Hawking flashes or via gravitational lensing
or they can cause clumping of dark matter. However, the shock waves are
very weak and speculations on resulting strong turbulences are on less
firm grounds.

\subsection{Baryon concentration}

\begin{figure}
 \vskip .01cm
 \center
 \epsfxsize=.41 \hsize \epsffile{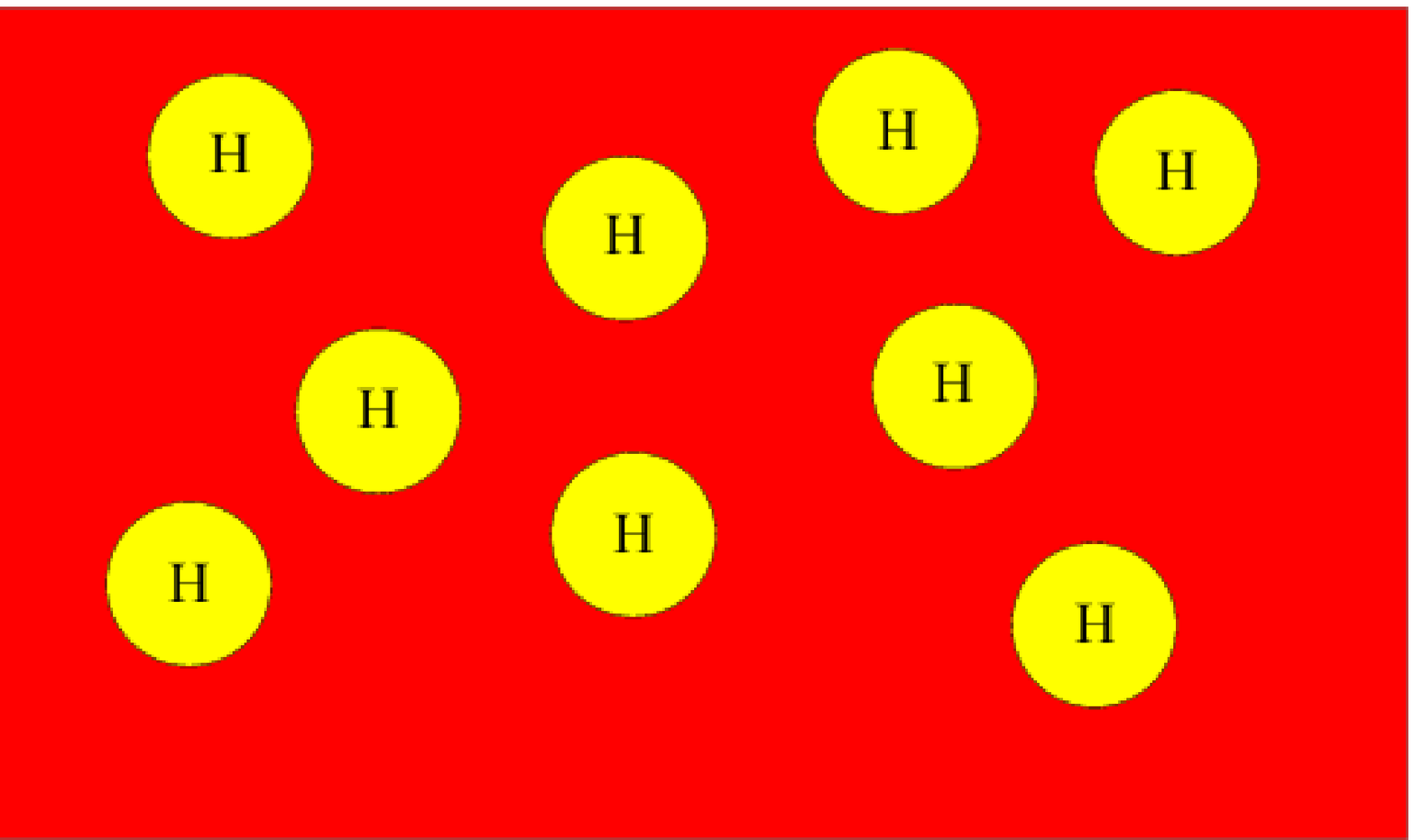}
 \hspace*{1cm}
 \epsfxsize=.41 \hsize \epsffile{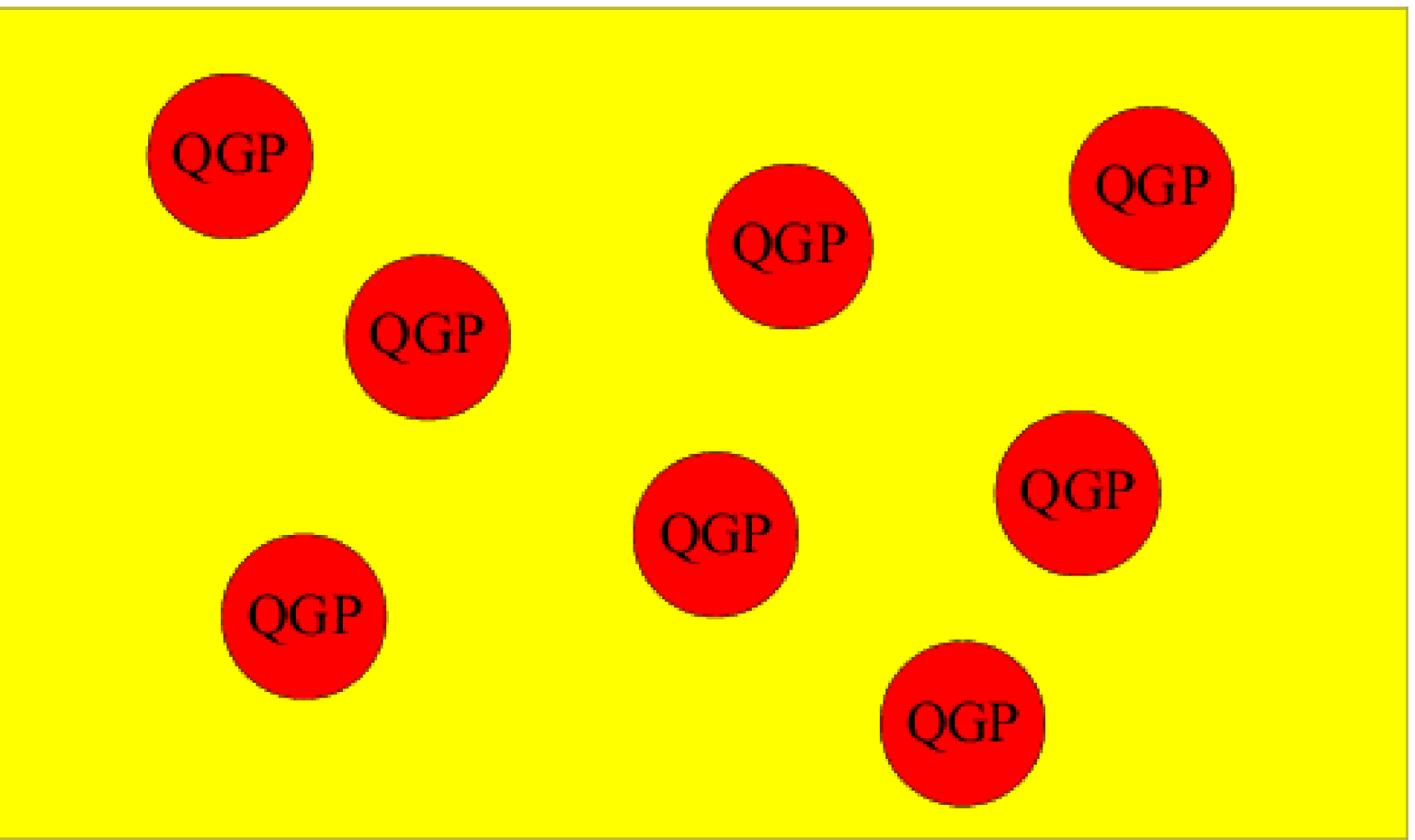}\\[3mm]
 \epsfxsize=.3 \hsize \epsffile{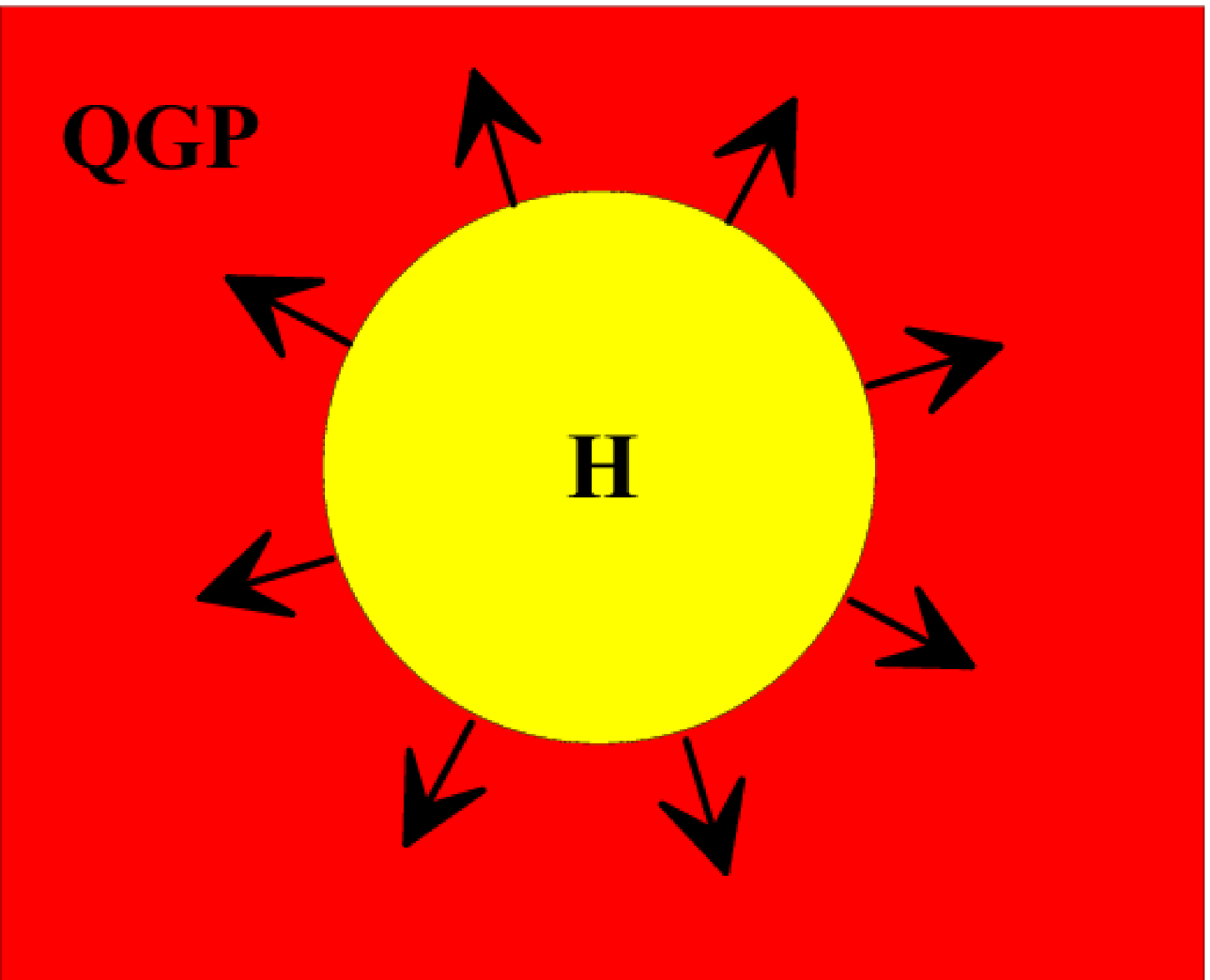}
 \hspace*{2cm}
 \epsfxsize=.3 \hsize \epsffile{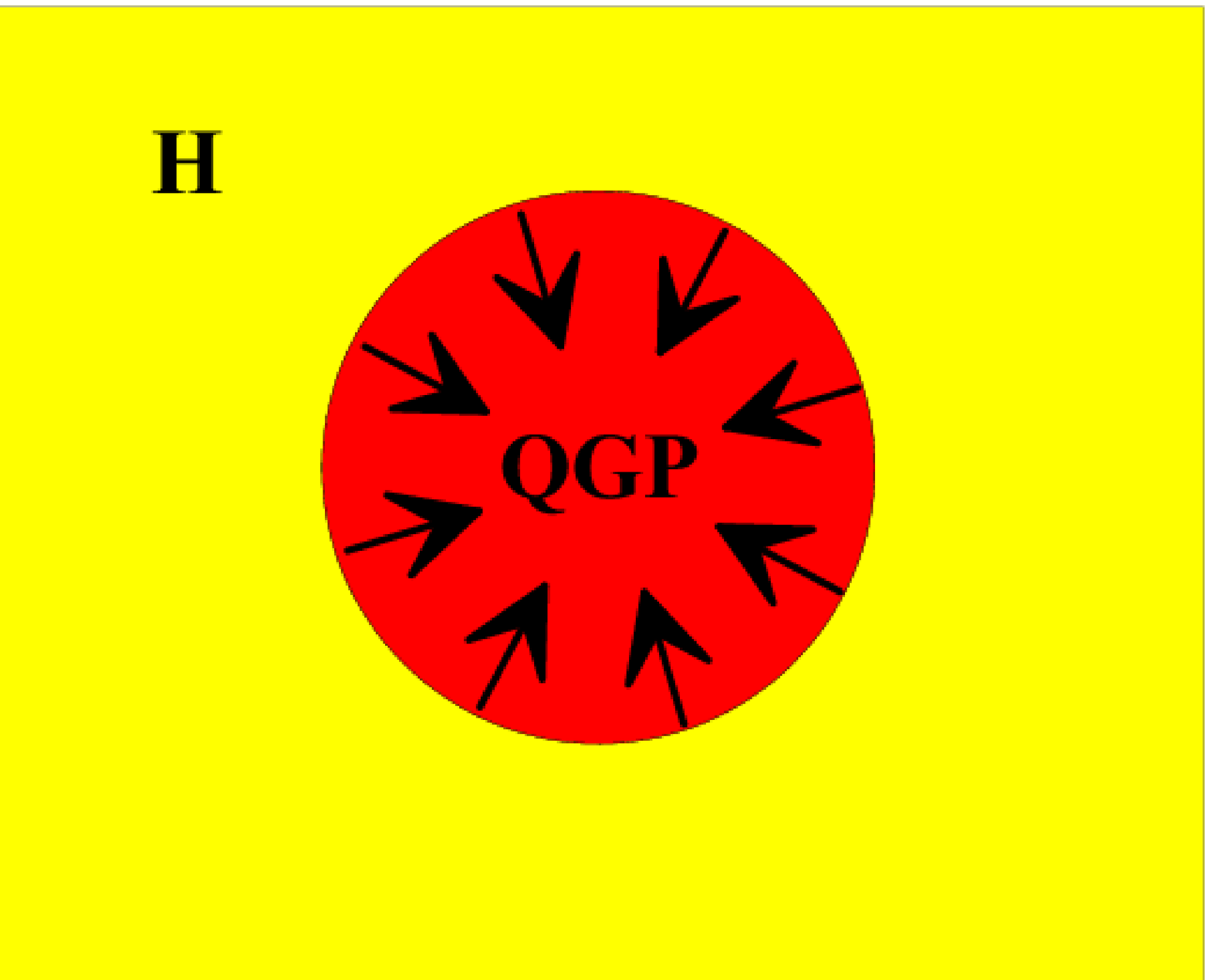}
 \caption{Left column: Growing bubbles of hadronic matter
 immersed in quark-gluon plasma.
 Right column: Shrinking islands of quark-plasma
 immersed in hadronic matter. For $T \ll 200$ MeV, the baryon charge is
 accumulated in the quark islands.}
 \label{islands}
\end{figure}

Let us now discuss a few possible relics of the cosmic confinement transition.
On small scales in the order of growing hadron bubbles
or vanishing last islands of the quark-gluon plasma (cf.\ Fig.~\ref{islands})
a baryon concentration arises.
According to the scenario of a near-to-equilibrium
transition, at the phase boundary also chemical equilibrium is maintained
between quarks (q) and nucleons (N).
The reaction $3 q \leftrightarrow N$ is unhindered and, due to detailed
balance, causes the balance equation for the baryo-chemical potentials
$3 \mu_q = \mu_N$. Since in the deconfined phase  the baryon charge is carried
mainly by the light current quarks, while in the confined phase the baryon
charge is in the very heavy nucleons (compared to the temperature scale),
the relation $\hat n^B_{QGP} \gg \hat n^B_H$ emerges from the
general expression of the net baryon density $\hat n^B$
(density of baryons minus density of anti-baryons)
in ideal gas approximation
(cf.\ (\ref{Euler}, \ref{pressure}))
\begin{eqnarray}
\hat n^B \approx  \left\{
\begin{array}{l}
\frac 13 N_f T^3
\left( \frac{\mu_q}{T} \right) \left[ 1 + \frac{1}{\pi^2}
\left( \frac{\mu_q}{T} \right)^2 \right] = \hat n^B_{\rm QGP},\\[3mm]
\frac{1}{\sqrt{2 \pi^3}} d_N T^3 \left( \frac{m_N}{T}\right)^{3/2}
\mbox{e}^{- m_N / T} \, \frac{\mu_N}{T} = \hat n^B_H
\label{nucleon_density},
\end{array}
\right.
\end{eqnarray}
where $N_f$ is the number of excited quark flavors (2 or 3 for
$u, d$ or $u, d, s$ quarks with negligible masses)
and $d_N = 4$ is the nucleon degeneracy;
$m_N = 939$ MeV denotes the averaged nucleon mass.
(\ref{nucleon_density}) is for $\mu_N < m_N$ 
and in Boltzmann approximation.
Due to the exponential suppression factor in (\ref{nucleon_density}),
the ratio $\hat n^B_{\rm QGP} / \hat n^B_H$ at $T \ll 200$ MeV becomes
very large.
In fact, the baryon charge is concentrated in the last islands of deconfined
matter, as visualized in Fig.~\ref{islands}.

This baryon concentration may cause two interesting effects intensively
debated during the last decade:
either the quark islands stabilize themselves and survive as
quark nugget relics or they convert into
isothermal baryon density fluctuations. We will discuss the former effect
in the next subsection. The latter effect can affect the primordial
nucleosynthesis if its survives up to then.
A necessary condition is
$l \sim 1 \, \mbox{m}\vert_{T = 100 \, MeV}$,
where $l$ is the comoving scale characterizing the mean distance of the
inhomogeneities, which is determined by the nucleation process
via $l \sim 0.8 (T_c / 100 \, \mbox{MeV}) \, d_{\rm nuc}$ \cite{Madsen}
with
\begin{eqnarray}
d_{\rm nuc} \approx 13
\left(\frac{145 \mbox{MeV}}{B^{1/4}} \right)^2
\left( \frac{4 B}{L} \right)
\left( \frac{\sigma}{0.035 B^{3/4}} \right)^{3/2} \,
\mbox{cm};
\end{eqnarray}
$L$ is the latent heat in the bag model (\ref{bag_eos}, \ref{pion_eos}),
cf.\ left upper panel in Fig.~\ref{eos_thermodynamics}.

In confined matter the neutrons can rapidly diffuse
thus causing an inhomogeneous neutron-to-proton ratio.
Recent re-analyses \cite{Madsen,Mathews} let this scenario appear less
probable: the low surface tension, discussed in subsection
\ref{small_supercooling},
gives rise to a typical scale
$l \sim 1$ cm and the detailed study of the nucleosynthesis
shows no deviations from the homogeneous standard scenario.

It should be mentioned, however, that the estimates
quoted above rely on the
homogeneous nucleation theory. Nucleation by impurities is more
efficient and, as shown in \cite{Madsen}, can indeed render
the inhomogeneity scale to $l \sim 1$ m, supposed there are pre-fabricated
fluctuations or impurities on an appropriate scale, such as vortices,
domain walls, magnetic monopoles, cosmic strings, or relic fluctuations from
the electroweak transition.
The required scale of $l \sim 1$ m corresponds just to the red-shifted
horizon scale at the electroweak transition (see below).
In \cite{Ignatius_Schwarz} it is pointed out that temperature fluctuations,
compatible with the COBE measurements, can give rise to a inhomogeneous
nucleation with interesting consequences.

Another idea is that a sufficiently large baryon concentration
can survive up to now in form of quark nuggets being thus candidates
for dark matter. For a recent discussion cf.\ \cite{Sinha}.
The quark nuggets must have baryon charges of $N^B \sim 10^{40 \cdots 45}$
because otherwise the baryon contrast would have diffused away.

\subsection{Strange matter}

Strangeness may play a particular role in the cosmic evolution.
While nowadays the hadrons are mainly nucleons with the
valence quark structure
$p = (uud)$, $n = (ddu)$, just after confinement also pions and kaons
were present. Kaons carry strangeness, e.g. $K^+ = (u \bar s)$,
but once the universe cools below 100 MeV they rapidly decay,
and the strange particles disappear from the world. They can be
afterwards newly created for a short while by strong interaction
of high-energy cosmic rays with matter and in the laboratory.
But the earlier universe was quite strange.
Due to the long expansion and cooling time scales in comparison with the
time scale of the weak interaction, there was chemical equilibrium
according to the reactions changing strangeness
\begin{eqnarray}
d \leftrightarrow u + e^- + \bar \nu_e, \quad
s \leftrightarrow u + e^- + \bar \nu_e.
\label{beta_equilibrium}
\end{eqnarray}
This results in the balance equations
\begin{eqnarray}
\mu_d = \mu_u + \mu_{e^-} - \mu_{\nu_e}, \quad
\mu_s = \mu_u + \mu_{e^-} - \mu_{\nu_e},
\end{eqnarray}
from which $\mu_s = \mu_d$ follows, or $n_s \approx n_d$.
Thus, a substantial part of the quarks were strange ones
(for a more detailed consideration including
charge neutrality cf. \cite{our_booklet}).
As pointed in \cite{Farhi,Witten}, within the bag model the strange
quarks can stabilize the quark matter. This idea is re-analyzed
with increasing sophistication \cite{strangelets} and there seems to be
a region in the parameter space still leaving the possibility
of stable strange quark nuggets. These represent candidates for dark matter.

The dedicated experiments with heavy-ion collisions, however, did not show
positive signals of strangelets \cite{no_strangelets}.
This constrains the parameter space
further and makes the strangelet idea less probable. It should be stressed,
however,
that in heavy-ion collisions the time scales are much shorter
and the conditions for creating strangelets may be less favorable.

\subsection{Effects of vanishing sound velocity}

As seen in Fig.~\ref{eos_thermodynamics} the sound velocity vanishes
in first-order
phase transitions in the mixed phase. This observation has an
important effect on the evolution of fluctuations: the restoring
pressure gradient of density-enhanced regions vanishes, and these regions
can collapse in free fall.

\subsubsection{Intermediate scales}

On intermediate scales, $l \ll \lambda \ll R_H$, the fate of the fluctuations
is followed by the cosmological perturbation formalism \cite{Schwarz}.
In particular, as pointed out in \cite{Schwarz},
kinetically decoupled cold dark matter can be trapped in
gravitational wells. Candidates for such dark matter are
WIMPzillas\footnote{This nickname refers to the notion of
\underline{W}eakly \underline{I}nteracting \underline{M}assive
\underline{P}articles, see \protect\cite{wimpzillas}.}
($M > 10^6$ GeV), primordial black holes
($10^{-18} \, M_\odot \ll M < M_\odot$), and axions ($M \sim 10^{-5}$ eV)
which can be accumulated to lumps with mass of
$M \sim 10^{-(10 \cdots 20)} M_\odot$. Also primordial gravitational waves
are modified, however, on scales today not accessible via pulsar timing.
Afterwards, the neutrino diffusion
is efficient enough to damp away the created density fluctuations
till nucleosynthesis, unless the decoupled matter lumps.

\subsubsection{Horizon scales}

As pointed out in \cite{Jedamzik}, on large scales,
$\lambda \sim R_H$, already previous causally non-connected weak
fluctuations on super-horizon scales can, after entry into the
horizon, collapse to black holes with
$M \sim 1 M_\odot$ due to $c_s^2 = 0$ during confinement
(cf.\ \cite{Schwarz} for a further discussion).
Such a mechanism can produce MACHOs\footnote{
MACHO is an acronym for
\underline{M}assive \underline{C}ompact \underline{H}alo
\underline{O}bjects, see \protect\cite{macho}.}
visible by gravitational lensing.

\subsubsection{Vanishing sound velocity during confinement?}

Since both effects are related to the vanishing of the sound velocity,
one has to analyze whether this may really happen. In the
case of a strong crossover, instead of first-order phase
transition, the sound velocity decreases but does not vanish completely.
As shown in
\cite{Schwarz} under such circumstances the resulting gravitational wells
are too weak to substantially capture dark matter.

One has also to mention that $c_s^2 = 0$ holds only for $p = const$
in the phase mixture region when all chemical potentials
$\mu_\alpha$ vanish or only one independent chemical potential occurs, i.e.,
$\alpha =1$. In cosmic matter there are numerous potentials, such as
$\mu_e, \mu_\mu, \mu_{\nu 's}, \mu_u, \mu_d, \mu_p, \mu_n$.

\begin{figure}
 \vskip .01cm
 \center
 \psfig{file=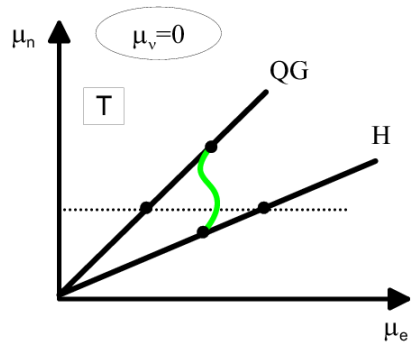,width=5.3cm,angle=-0}
 \caption{Constructing the phase transition in presence of a some chemical
 potentials. Imposing charge neutrality yields the two curves
 labeled by QG and H
 for deconfined and confined matter, respectively. Pressure equality at
 fixed temperature and at equal neutron-baryon chemical potential $\mu_n$
 (indicated by the dotted line) is connected
 with a jump of the electron chemical potential $\mu_e$. Instead, averaged
 charge neutrality results in the gray S-shaped line. On this line,
 the pressures of both phases are equal but they change when moving along
 this line.}
 \label{sound_velocity}
\end{figure}

Imposing $\beta$ equilibrium, according to (\ref{beta_equilibrium})
and $n \leftrightarrow p + e^- \bar \nu_e$,
and local electric charge neutrality,
at least one of the chemical potentials changes
discontinuously in the case of a naively constructed phase transition.
Instead, as stressed in \cite{Glendenning}, constructing the phase
equilibrium with the averaged charge neutrality
$n_{had} + n_{QGP} = 0$ results
(cf.\ Fig.~\ref{sound_velocity})
in $p \ne const$ in the mixed state, and
therefore the sound velocity does not strictly vanish.
The effect of the various chemical potential may be small if
$\mu_\alpha \ll T$ and the sound velocity at least substantially drops
in the mixed state.

\begin{figure}
 \vskip -.01cm
 \center
 \psfig{file=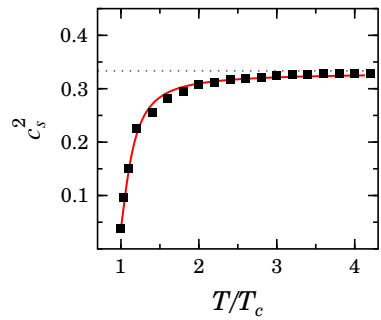,width=5.9cm,angle=-0}
 \caption{The sound velocity of gluon matter as a function of
 temperature. Symbols depict the results of lattice QCD data from the
 group Bielefeld
 and the curve results from the quasi-particle model
 \cite{our_QCD_paper}.}
 \label{c_s2}
\end{figure}

Otherwise one should notice that upon approaching the confinement temperature
the sound velocity in the deconfined state itself already drops
(cf.\ Fig.~\ref{c_s2}). This is usually related to a softening of the
equation of state. In \cite{Danielewicz} evidences of such a softening
is extracted from data of heavy-ion collisions at AGS energies.
It is matter of debate whether
this is an indication of deconfinement effects at such comparatively low
beam energies of 10 $\cdots$ 15 A$\cdot$GeV.

\begin{figure}
 \vskip -.01cm
 \center
 \psfig{file=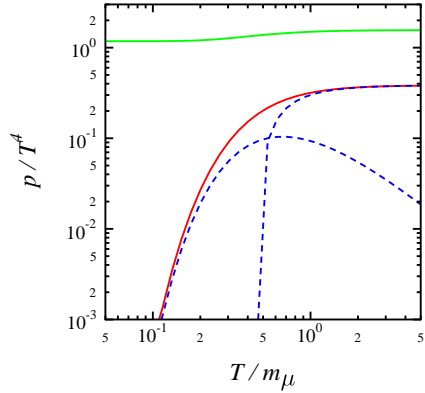,width=5.9cm,angle=-0}
 \caption{The scaled pressure as a function of the temperature
 of the muons (lower solid line: exact evaluation of the
 statistical integral (\ref{pressure}),
 dashed lines: using the first two terms
 of the high and low temperature expansion). Upper gray curve: total pressure
 taking into account the background of electrons, photons, and three
 neutrino species.}
 \label{pressures}
\end{figure}

Moreover, the annihilation process of certain particle species
is also related to a decrease of the sound velocity,
cf.\ \cite{Schwarz}.
This turns out to be a marginal effect. To look into
details let us consider the era when the muons disappear.
At $T \gg m_\mu$ and at $T \ll m_\mu$ the
degeneracies are
$d_{\rm eff} =$ 14.25 and 10.75, respectively.
By evaluating numerically the statistical integral (\ref{p_statistic}),
one gets the pressure as displayed in Fig.~\ref{pressures}.
Indeed, there is a small change of the scaled total pressure at
$T/m_\mu \sim 0.4$. At $T/m_\mu < 0.2$, the muons can be neglected and at
$T/m_\mu > 1$ the radiation approximation,
$p_\mu = \frac{d_\mu \pi^2}{90} T^4$ with $d_\mu = 4$, is sufficient.

\begin{figure}
 \vskip -.01cm
 \center
 \psfig{file=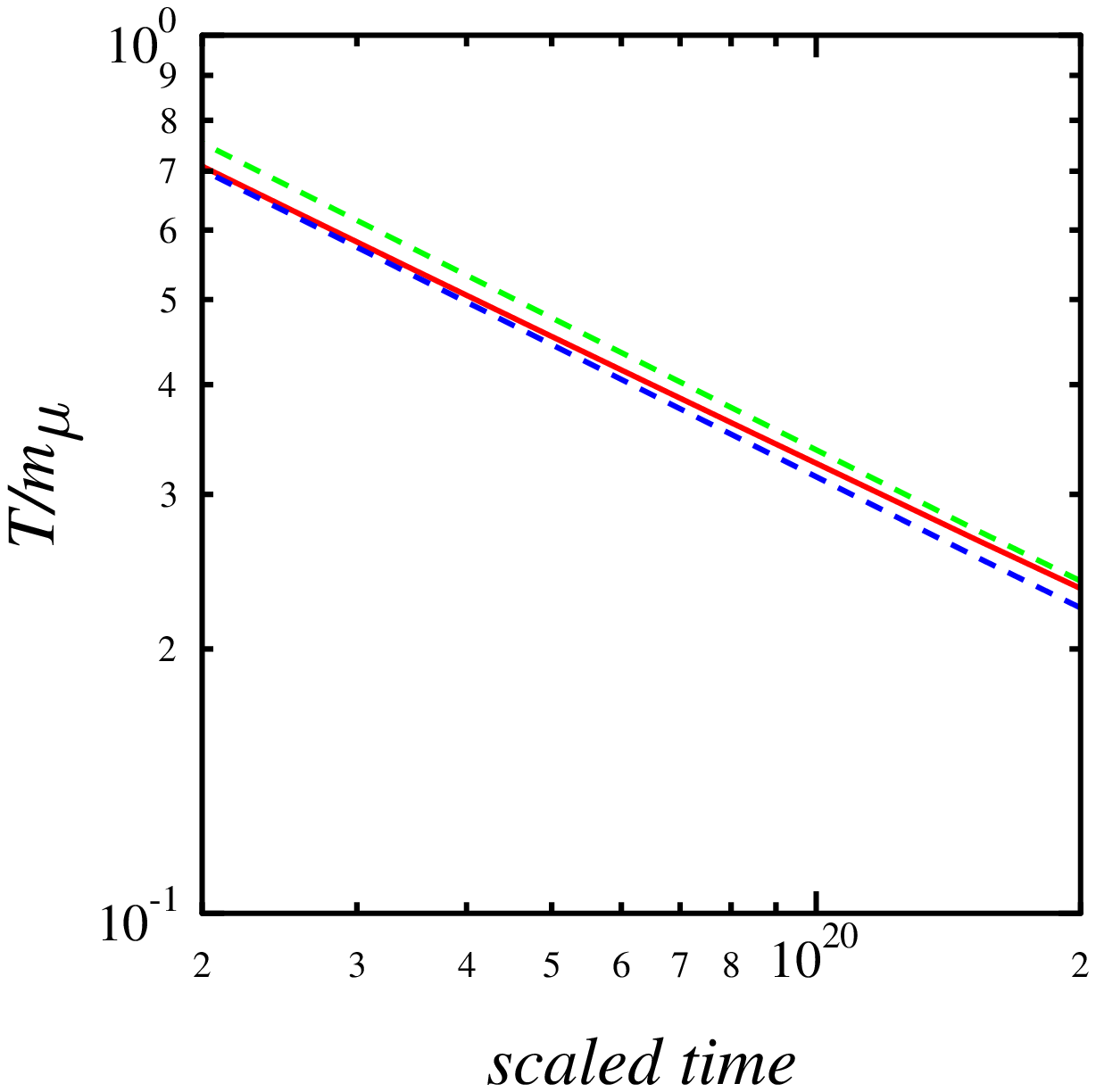 ,width=5.6cm,angle=-0}
 \hfill
 \psfig{file=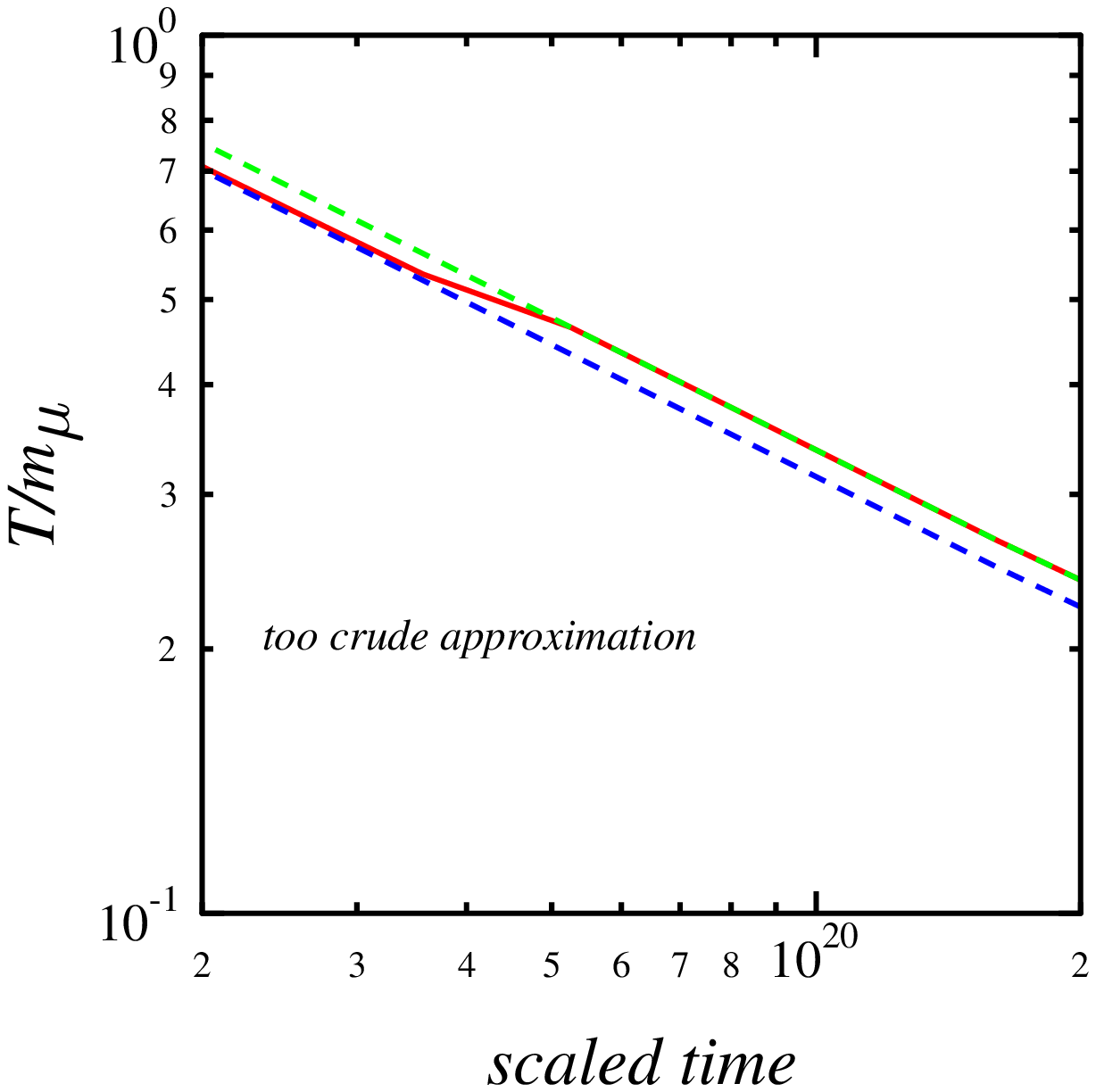 ,width=5.6cm,angle=-0}
 \caption{The evolution of the temperature as a function of the scaled
 time. Lower dashed line: only the background, upper dashed line: background
 plus massless muons, solid line: background plus muon contribution.
 The left panel uses the correct evaluation of the pressure, while the
 right panel employs only the expansion described in text.}
 \label{T_mu}
\end{figure}

Solving the evolution equations (\ref{Friedmann1}, \ref{Friedmann2})
results in the temperature
history as displayed in the left panel of Fig.~\ref{T_mu}.
There is a very moderate change of the cooling due to muon annihilation.
Only when incorrectly approximating the muon pressure by the first
two terms in the high-temperature expansions one gets a more pronounced
effect which looks like a phase transition
(see right panel of Fig.~\ref{T_mu}).

Thus the effect of particle annihilation is weak and unlikely to cause
a reheating or a substantial softening of the equation of state.
This can also be seen from the measure
$3 p/e$ which drops from  1 to 0.95 during muon annihilation
and rises to 1 afterwards.
Therefore, particle annihilation can hardly cause strong effects,
as mentioned in \cite{Schwarz}.

\subsection{Interim resume}

%Let us make here a interim summary.
The confinement transition is quite interesting, but the cosmic evolution
is already slow compared with the strong interaction time scale,
and probably the most severe effects are washed out.
Quasi-equilibrium means memory loss.
In addition some details are poorly known, such as the correct order
of the transition and the surface tension (if any).
Therefore, one might argue that earlier stages with more rapid evolution
are out of equilibrium and do not leave such fragile imprints.

\section{Electroweak symmetry breaking}

In the electroweak standard model \`a la Glashow-Salam-Weinberg
the Higgs field is responsible for the dynamical mass generation via
spontaneous symmetry breaking. At sufficiently high temperatures,
$T > 100$ GeV, the expectation value of the Higgs field is zero,
i.e., the symmetry is restored and particles are massless.
At $T < 100$ GeV the symmetry becomes broken because
$\langle \mbox{Higgs} \rangle \ne$ 0
and particle masses become finite.
For Higgs masses $m_H < 75$ GeV,
the symmetry breaking appears as weak first-order
phase transition, while for the more likely case of
$m_H > 75$ GeV a phase transition of second order happens
\cite{electroweak_transition_1,electroweak_transition_2}.
Indeed, approaching $m_H \to 75$ GeV, the surface tension
drops dramatically \cite{electroweak_transition_2}.
In view of the experience with the confinement transition,
one may not expect drastic effects from this transition.

However, there is an important issue, outside the scope of the present
work, related to non-perturbative effects of solitary field configurations
called sphalerons:
baryon and lepton numbers are not conserved. Therefore, at the electroweak
transition the baryon and lepton excess of the universe might be
generated. Since a final consistent picture does not seem to be elaborated
yet, we refer to some recent surveys on this topic
including also energies of the GUT scale, see \cite{baryo_genesis}.

\section{Inflation \label{Inflation}}

Equation (\ref{R2dot}) shows that at $p < - \frac 13 e$ the expansion of the
universe becomes accelerated. In particular, if the vacuum energy density
dominates, i.e. $e^{\rm vac} \gg e_\alpha$ and
$ e= e^{\rm vac} = - p^{\rm vac} \approx const$
then $R \propto \mbox{e}^{{\cal C} \sqrt{e^{\rm vac}} t}$.
A series of problems in modern cosmology is solved
by inflation if the scale factor $R$ grows in the inflationary period
by a factor of about $10^{30}$.
These standard problems, like horizon, flatness, homogeneity, age, and
monopole problems, are summarized in many surveys, see
\cite{cosmology_book}.

Inflation can be realized within a model
with a dominating spatially homogeneous
scalar field which, via a potential $V(\phi)$, is self-interacting.
The corresponding equation of state can be deduced from
\begin{eqnarray}
e & = & +V(\phi,T) + \frac 12 \dot \phi^2,\\
p & = & -V(\phi,T) + \frac 12 \dot \phi^2
\end{eqnarray}
(quantum corrections cause a temperature dependence
of the effective potential).
For a sufficiently slow evolution, the kinetic term can be neglected
and the coherent field mimics a perfect vacuum.
Inflation is expected to happen on the
GUT scale of $T \sim 10^{16}$ GeV and provides,
via adiabatic Gaussian fluctuations,
the seed for later structure formation.
The various inflationary scenarios differ essentially only in the choice
of the potential. To recall a few scenarios (cf.\ \cite{inflation}):
\begin{itemize}
\item
old inflation: decay of a metastable (``wrong'') vacuum caused by a
potential with two local minima corresponding to a phase transition
of first order,
\item
new inflation: a potential which is very flat around
$\phi_i = 0$, where $\phi_i$ is the initial state,
and which has a minimum at $\phi_R > 0$,
\item
chaotic inflation: a simple U-shaped potential and
$\phi_i \gg 0$ evolving to $\phi_R = 0$,
\item
hybrid inflation: two coupled scalar fields with
$\varphi \gg \phi$, and $\phi$ evolves in the slow rolling down potential
of new inflation.
\end{itemize}
We do not consider here scenarios beyond Einstein's theory or involving
supersymmetry or string theory.

\subsection{Viscous inflation \label{viscosity}}

The original version of the old inflationary scenario \cite{Guth}
assumed a supercooling by 30 orders of magnitude to achieve the stretching
of $R$ by 30 order of magnitudes. Then, of course, the exponential
grow of the space prevents a graceful exit to a homogeneous state after reheating.
However, there are also phenomenological ans\"atze which realize
successful inflation without such tremendous supercooling.
For instance, in viscous inflation \cite{viscous_inflation}
a suitably tuned volume viscosity can compensate the cooling,
moreover, due to the replacement of the pure thermodynamic pressure $p$ by
$p - 3 \frac{\dot R}{R} \hat \xi$ (see (\ref{p_ij})),
the space expansion can accelerate  for suitable values of $\hat \xi$.
Then there happens a tremendous entropy production. To be specific,
an increase of the comoving entropy, $s \, R^3$, by a factor $10^{90}$
is sufficient to solve most of the standard problems.
The key here is entropy production, which spoils the comoving
entropy conservation and such an
information link of very early and late stages of the universe.

\subsection{Tepid inflation}

\begin{figure}
 \vskip -.01cm
 \psfig{file=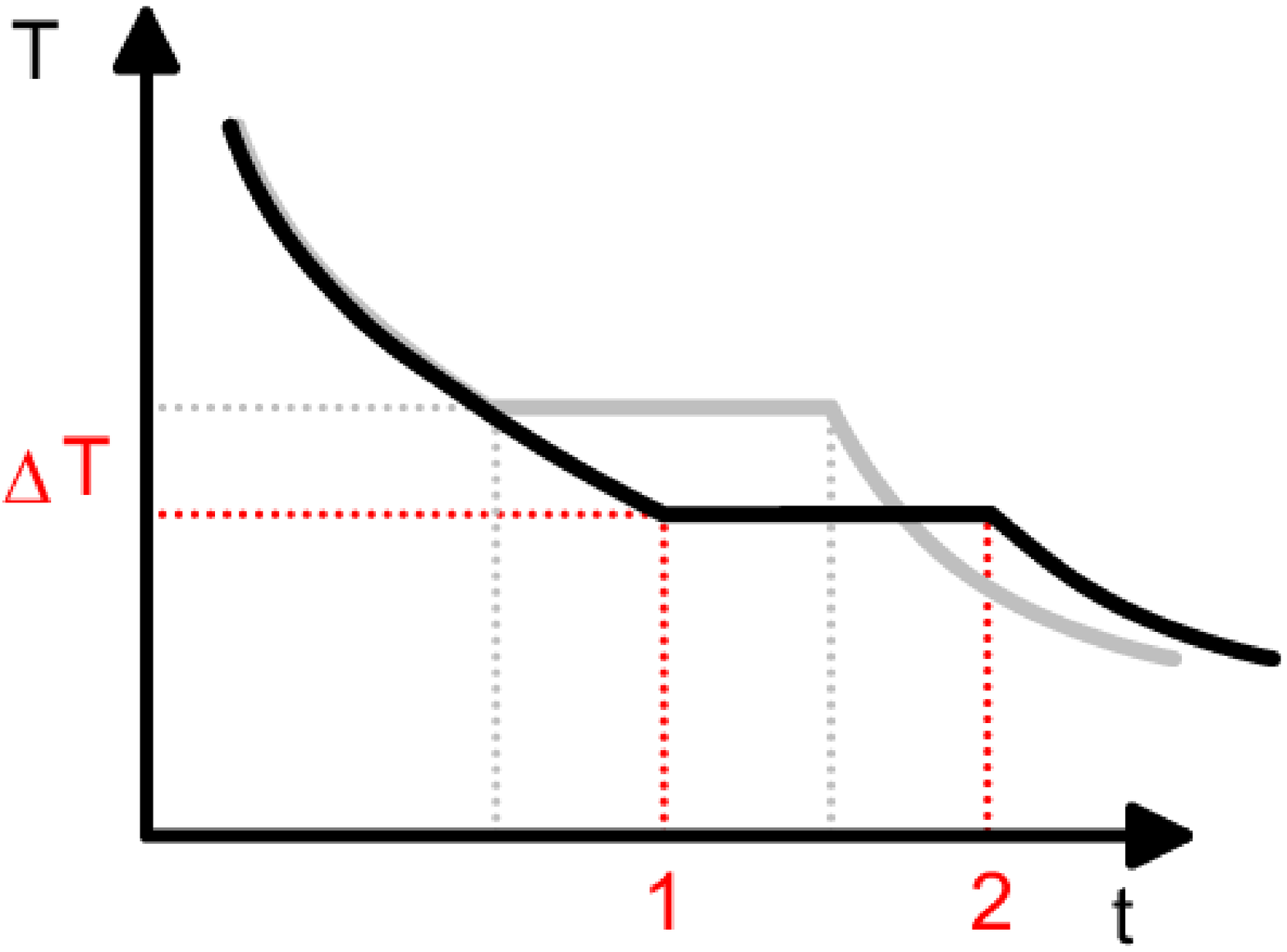,width=3.9cm,angle=-0}
 \hfill
 \psfig{file=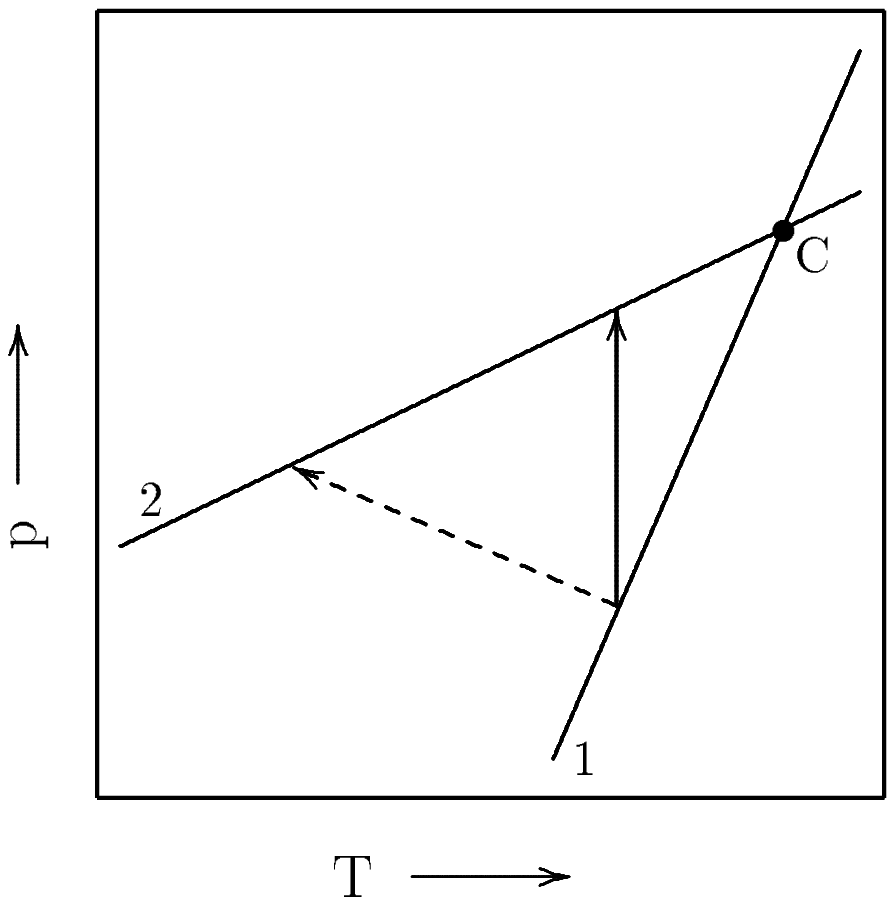,width=3.9cm,angle=-0}
 \hfill
 \psfig{file=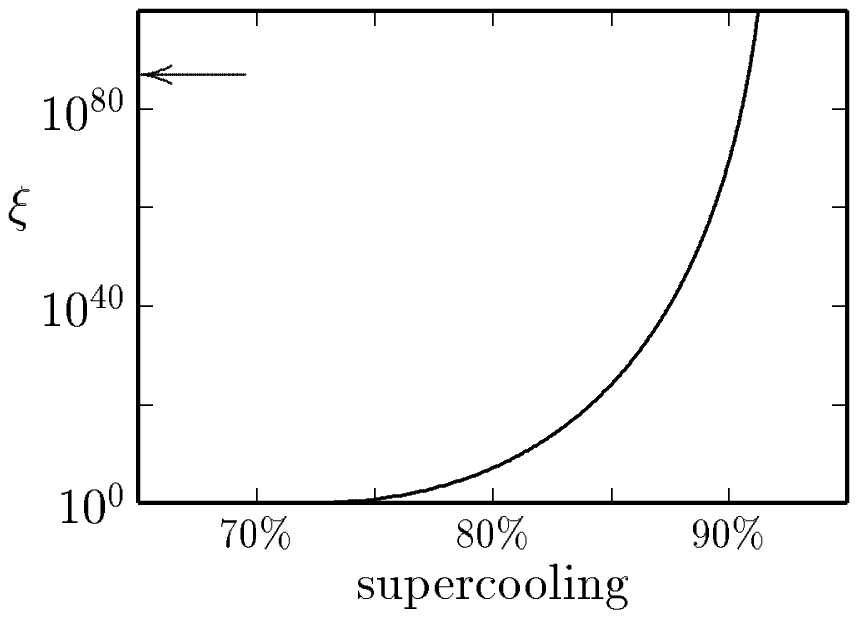,width=3.9cm,angle=-0}
 \caption{Left panel: Isothermal phase transition with supercooling
 $\Delta T$ (heavy solid curve).
 Middle panel: Sketch of the equation of state near the equilibrium
 point ``C''.
 Right panel: Entropy increase as a function of supercooling.}
 \label{isothermal_pt}
\end{figure}

Another way of entropy production is a phase transition with supercooling.
Let us consider as an example an isothermal phase transition at
$T = const$, as visualized in the left panel of Fig.~\ref{isothermal_pt}.
The state of matter changes as indicated in the middle panel by the
solid vertical arrow, i.e., there is a transition where the new phase (2)
has larger pressure than the old phase (1). This causes a compression of the
regions filled with the old phase. Taking into account
the averaging prescription for the energy density during the phase transition
\begin{eqnarray}
e = e_1 x + e_2 (1-x)
\end{eqnarray}
and analog equations for the pressure and the entropy density,
the entropy increase for the perfect 2-phase fluid is found
from (\ref{entropy_current}) as
\begin{eqnarray}
(s R^3)\dot{} & = & R^3 \left\{
\frac{2}{T_1 + T_2} (p_1 - p_2) \dot x  +
(T_1 - T_2) \left[ \dot \Delta + 3 \frac{\dot R}{R} \Delta \right] \right\},\\
\Delta & = & x s_1 - (1-x) s_2, \nonumber
\end{eqnarray}
instead of $(s R^3)\dot{} = 0$ for the ideal 1-component fluid.
To quantify the entropy production let us define
an entropy increase factor by $\xi = \frac{s R^3}{s_i R_i^3}$.
This factor is displayed in the right panel of Fig.~\ref{isothermal_pt}
as a function of the supercooling. One observes that moderate supercooling
in the order of 90 - 95\% is sufficient for successful inflation.

\begin{figure}
 \vskip -.01cm
 \center
 \psfig{file=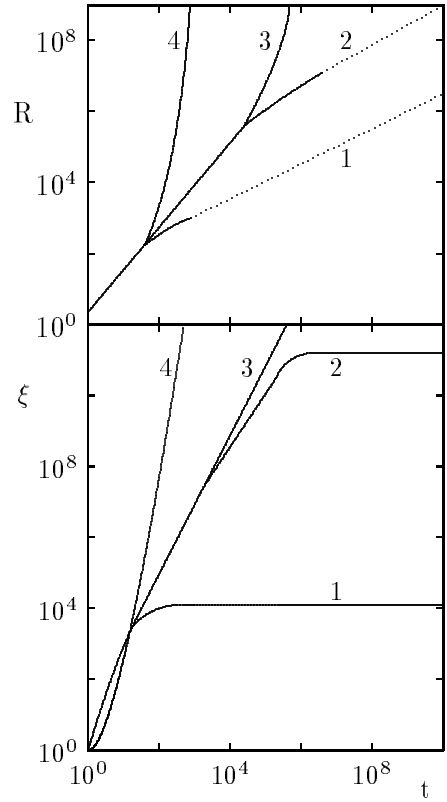,width=6cm,angle=-0}
 \caption{Evolution of the scale factor (upper panel) and the
 entropy increase factor (lower panel) as a function of time.
 The dotted sections in the upper panel correspond to the flat sections
 in the lower panel and indicate that the old phase disappeared.
Curves labeled by 1, 2, 3, 4 correspond to
$\sigma_0 =$ 2.064, 2.06499, 2.065, 2.066, respectively.
For more details, like the parameterization of the equation of state,
consult \protect\cite{our_booklet,my_inflation}.}
 \label{tepid_bubble_inflation}
\end{figure}

\begin{figure}
 \center
 \psfig{file=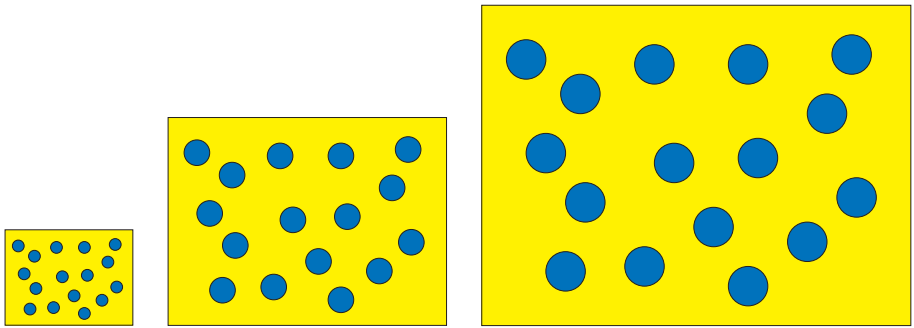,width=1.7cm,angle=-0} \quad
 \psfig{file=17_1.eps,width=3.7cm,angle=-0} \quad
 \psfig{file=17_1.eps,width=5.7cm,angle=-0}
 \caption{Visualization of the graceful exit problem:
 the space expands too rapidly and the growing bubbles fail to
 percolate since nucleation is, due to large
 supercooling, too rare to fill the space.}
 \label{expanding_space}
\end{figure}

In \cite{tepid_inflation} we have discussed a more dynamical
model with a phase transition and an ansatz for the relaxation
time approximation for $x_2(t)$. We find a sufficient
tepid inflation. Now we consider the nucleation theory, presented
in section \ref{nucleation_theory},
and study the evolution of the scale factor
and the entropy increase.
The results of a numerical integration of the evolution equations
(\ref{Friedmann1}, \ref{Friedmann2}, \ref{w_0}, \ref{Avrami})
are presented in Fig.~\ref{tepid_bubble_inflation}.
There are two regimes:\\
(i) a too small surface tension parameter $\sigma_0$ causes a too
small supercooling and insufficient inflation, while\\
(ii) a too large value of $\sigma_0$ repeats the graceful exit problem
of old inflation, i.e. the space is exponentially growing, and the growth
of bubbles
is too slow to fill the space, as visualized in Fig.~\ref{expanding_space}.

However there is a small corridor in parameter space where we find
power law inflation with
$R \sim t^\alpha$, $\alpha > 1$,
resembling the extended inflation in Brans-Dicke theory.
Numerically it is difficult to follow a successful inflationary evolution
due to the needed fine tuning of $\sigma_0$ and the accuracy of the
integration procedure with respect to the large number problems.
Therefore, it is hard to make statements on the emerging fluctuation
spectrum.

\section{Summary}

We should be aware that the evolution of the universe is accompanied
by a permanent change of the states of matter, which can be in some
interesting cases be considered as phase transitions.
There is a change of certain order parameters, like
a hypothetical scalar field $\phi$ on the GUT scale, the
Higgs field on the electroweak scale,
the chiral condensate on the QCD scale, or even now the quintessence.
This causes partially drastic changes of matter like the
transition from freely roaming quarks and gluons to hadrons,
or the binding of nucleons into light nuclei during nucleosynthesis,
or simply the annihilation of particles like $e^+ e^- \to 2 \gamma$.
These changes induce also changes of the expansion dynamics of the universe,
most drastic during inflation, e.g..
Furthermore in the transition from a radiation dominated
universe to a matter dominated one, the time-evolution of the scale
factor $R(t)$ changes from $ R \propto t^{1/2}$ to $R(t) \propto t^{2/3}$.
Some uncertainty is introduced by the evidence recently found
of a dominating vacuum energy density, the quintessence, because its
back-extrapolation needs more definite knowledge on its nature.

Assuming that in early stages quintessence is not dominating the
energy density, we considered in some detail the
confinement transition in the Big Bang and, very briefly, contrasted it
to the Little Bang, presently investigated in heavy-ion collision experiments.
In the Little Bang one observes direct
electromagnetic radiation like real photons and virtual photons via
$\gamma^* \to e^+ e^-$. This points to temperature scales of
170 MeV and larger, i.e., above the deconfinement transition.
In contrast to this,
there do not seem to be specific definite relics from
the confinement era of the Big Bang, apart the protons and neutrons
created in hadrosynthesis.
We summarized possible relics, whose existence is hypothetical and the
predictions of which are sometimes related to poorly known details of the
confinement dynamics.

Whereas the results of the nucleosynthesis processes specifically depend on
the interplay of neutrino decoupling, neutron life time, and
cosmic cooling rate, the hadrosynthesis at confinement seems to be rather
unspecific. Nucleons emerge as other hadrons do as well, but the unstable
hadrons decay shortly after confinement and the proton-to-neutron ratio
is determined by $\beta$ equilibrium. Hence, hadrosynthesis does not cause
a specific hadron composition due to the cosmic evolution
since chemical equilibrium means memory loss.

Strictly speaking, the success of the primordial nucleosynthesis
makes us strongly believe in a maximum temperature of
$T_R > 1$ MeV, while the prediction of a flat universe
and of an appropriate fluctuation spectrum for
structure formation by inflationary scenarios makes us believe
in $T_R > 10^{15}$ GeV. Within these scales, a consistent picture
of the evolution of the universe seems to emerge, but it
leaves a consistent treatment of the ``initial state'' and
the reheating (or, better, ``pre-heating'')
to $T_R$ as the most challenging problem probably to be solved
within quantum gravity.

%%%%%%%%%%%%%%%%%%%%%% Acknowledgements %%%%%%%%%%%%%%%%%%%%%%%%%%%%%%%%%%
\vspace*{0.25cm} \baselineskip=10pt{\small \noindent
Acknowledgments:
K. Gallmeister is grateful for preparing
the figures. I thank
F.W. Hehl and A. Gross for carefully reading
the manuscript.
Some of the presented material is an up-grade of
\cite{our_booklet}. The author thanks his colleagues for the fruitful
collaboration. The work is supported in part by BMBF grant 06DR829/1.}

%%%%%%%%%%%%%%%%%%%%%%%%% Bibliography %%%%%%%%%%%%%%%%%%%%%%%%%%%%%%%%%%%
% \bibitem{journal} Author1, Author2, and Author3, Journal
%   {\bf Volume} (Year) Page_number
% \bibitem{book} Author1, Author2, and Author3, {\it Book_Title},
%   Publisher, Place Year
% \bibitem{edited} Author1 and Author2, in {\it Book_Title} edited by
%   Author3 and Author4, Publisher, Place Year, p. Page_Number
% \bibitem{thesis} Author, PhD thesis, University, Place Year
% \bibitem{eprint} Author1, Author2, and Author3, cond-mat/9876543,
%   submitted to Journal
%%%%% OR
%   accepted for publication in Journal
%%%%%%%%%%%%%%%%%%%%%%%%%%%%%%%%%%%%%%%%%%%%%%%%%%%%%%%%%%%%%%%%%%%%%%%%%%

\end{document}